\documentstyle[preprint,eqsecnum,aps]{revtex}

\newtheorem{theorem}{Theorem}
\newtheorem{definition}{Definition}
\tighten

\begin{document}
\draft
\title{Classical electrodynamics of point charges}
\author{Massimo Marino\thanks{Email address: massimo.marino@mi.infm.it}}
\address{INFM - Dipartimento di Fisica, Universit\`{a} di Milano, \\
via Celoria 16, I-20133 Milano, Italy}
\maketitle

\begin{abstract}
A simple mathematical procedure is introduced which allows
redefining in an exact way divergent integrals and limits that
appear in the basic equations of classical electrodynamics with
point charges. In this way all divergences are at once removed
without affecting the locality and the relativistic covariance of
the theory, and with no need for mass renormalization. The
procedure is first used to obtain a finite expression for the
electromagnetic energy-momentum of the system. We show that the
relativistic Lorentz-Dirac equation can be deduced from the
conservation of this electromagnetic energy-momentum plus the
usual mechanical term. Then we derive a finite lagrangian, which
depends on the particle variables and on the actual
electromagnetic potentials at a given time. From this
lagrangian the equations of motion of both particles and fields
can be derived via Hamilton's variational principle. The
hamiltonian formulation of the theory can be obtained in a
straightforward way. This leads to an interesting comparison
between the resulting divergence-free expression of the
hamiltonian functional and the standard renormalization rules for
perturbative quantum electrodynamics.
\end{abstract}

\pacs{03.50.De, 11.10.Ef}

\section{Introduction}

The relativistic Lorentz-Dirac equation was derived in 1938
\cite{uno} to describe the motion of a classical point charge in
an arbitrarily assigned external field, taking into account its
electromagnetic self-interaction and therefore the effect of
radiation reaction (useful reviews on this subject can be found in
Ref. \cite{tre,treb}).

Dirac believed that, once a finite classical theory for point-like
charges had been obtained, then by applying appropriate rules of
canonical quantization it should be possible to formulate the
quantum theory in a way that avoided the occurrence of troublesome
infinities. Already in the final part of his original paper
\cite{uno}, with the explicit purpose of suggesting the path
towards a possible quantization, Dirac showed that his theory
could be deduced from a variational principle, by making use of a
non-local action which had been previously proposed by Fokker
\cite{trec} to describe a system of charged particles interacting
at a distance through the average of the retarded and advanced
fields. However Dirac soon realized that such an action could not
allow to reexpress the dynamics in hamiltonian form. For this
reason, soon thereafter he produced a second paper \cite{due}
specifically devoted to the problem of achieving a canonical
quantization of his theory. In this work he abandoned the action
principle, and attempted to write down directly a four-dimensional
hamiltonian in terms of an unphysical field dependent on
the proper time of each particle constituting the system under
consideration. For this field Dirac skilfully postulated particular
{\it ad hoc} Poisson brackets, involving an arbitrary time-like
constant four-vector $\lambda$, in such a way that his
covariant equation of motion for the particles was recovered in
the limit of vanishing $\lambda$. It is not surprising that such
an involute approach could hardly lend itself to further
generalizations, and Dirac was unable to use it as the starting
point for a consistent formulation of quantum electrodynamics. It
is also interesting to observe that, although Dirac had derived in
\cite{uno} his equation of motion using essentially the principle
of local energy-momentum conservation, he did not provide a
general expression for the global conserved quantities. Besides,
the conservation laws no longer emerged in an obvious way from the
canonical formalism that he proposed.

With the advent of renormalization theory, perturbative quantum
electrodynamics developed in an impressive way without relying on
any input from classical theory for the formal treatment of
divergences. However, research on the classical theory was never
abandoned. In particular, a new finite classical action describing
the behavior of both particles and fields was obtained by Rohrlich
\cite{tre,quattro} by adding an appropriate term to the original
expression of Fokker and Dirac. However, this result could not
help in providing a canonical formulation of the theory.
Furthermore, in the action the electromagnetic field was split as
the sum of a source-free term plus the contribution generated by
the particles through the retarded and advanced potentials. This
contrasted with the lagrangian formulation of QED which, despite
the subtleties related to the problem of providing an adequate
renormalization, is always expressed only in terms of the full
interacting electromagnetic field. Similar considerations apply
also for other studies subsequently appeared, in which the
expression of the electromagnetic energy-momentum in terms of the
retarded fields was subjected to an accurate analysis
\cite{cinque}, and a different kind of lagrangian, based on the
distribution theory, was derived \cite{sei}.

In the present paper we shall propose a simple and general method
to evaluate the integrals of the electromagnetic energy-momentum,
lagrangian and hamiltonian densities over spatial regions
containing the point particles. In a similar way, we shall assign
a finite value to the fields and potentials on the particles'
world-lines, by generalizing a procedure previously proposed by
Teitelboim and Lopez \cite{sette,otto}. In this way practically
all the familiar formulae of electrodynamics, which are commonly
applied when the particles' self-interaction is neglected, can be
immediately generalized to the self-interacting case by
appropriately reinterpreting each individual term according to the
given prescriptions. The resulting equations of motion will be the
relativistic Lorentz-Dirac equation for the particles and the
Maxwell equations with point-like sources for the fields. In
particular we shall obtain:
\begin{enumerate}
\item  A finite expression for the total energy-momentum of the
system of fields and particles, which is Lorentz-covariant and
conserved on the solutions of the equations of motion;
\item  A finite lagrangian which determines the dynamics of both
fields and particles according to Hamilton's variational
principle;
\item  A finite hamiltonian and a system of Poisson's brackets which
express the dynamics in canonical form.
\end{enumerate}

In the treatment here presented, the infinities will be
consistently removed without altering any of the fundamental
formal properties of the theory, including locality and
relativistic covariance. A remarkable property of this procedure
is that it allows expressing the four-momentum, the lagrangian and
the hamiltonian of the system only in terms of the particle
variables and of the total instantaneous electromagnetic field,
i.e.\ without any unphysical splitting into retarded, advanced,
incoming or outcoming parts. This will provide a formulation of
the theory which is in complete agreement with the requirements of
local relativistic field-theory. Furthermore, no diverging
parameter will have to be introduced, since only the physical mass
of the particles will be used. Due to the close correspondence
with the quantum theory, we shall show that these results may also
suggest a new way of formulating the renormalization procedure of
quantum electrodynamics.

The paper is organized as follows.
In Sec.\ II we define the renormalized energy-momentum associated with
the electromagnetic field in the presence of point charges, and we
discuss how the conservation of the sum of this energy-momentum and the
familiar mechanical term associated with the particles (with mass parameter
equal to the observable mass $m$) determines the equation of motion of the
particles, provided that the field satisfies the usual Maxwell equations.
Actually, this provides a new rigorous and straightforward derivation of the
relativistic Lorentz-Dirac equation. We shall comment on the mathematical
structure of the system of coupled equations of motion for particles and
fields, and on the non-trivial constraints that are to be imposed on the
initial data of the corresponding Cauchy problem. In Sec.\ III we
define the renormalized lagrangian of the system, and we show how the
equations of motion of both particles and fields can be derived from an
action principle. We shall see that on the solutions of the field equations
the value assumed by our lagrangian can be approximated by an expression
corresponding to an electron of finite radius $r$ and bare mass $%
m_{0}=m-q^{2}/8\pi r$, although the canonical forms of the two
theories remain essentially inequivalent even in the limit
$r\rightarrow 0$. In Sec.\ IV we give a precise mathematical
definition for the partial derivatives of the lagrangian with
respect to the particle variables, and we make use of this
definition to put the equations of motion in the standard form of
Euler-Lagrange equations. The reformulation of the theory in the
hamiltonian formalism will then be obtained in a straightforward
way via a Legendre transformation. In Sec.\ V we obtain an
equivalent expression for the renormalized hamiltonian in terms of
the Fourier transform of the electromagnetic potential, and in
Sec.\ VI we make use of this expression to illustrate the close
relationship with the renormalization rules for the electron
self-energy in perturbative quantum electrodynamics.

Most of the
mathematical proofs and technical developments are worked out in
the Appendixes. In particular, in App.\ A we consider the retarded
and advanced fields near a point charge in arbitrary motion, and
we show that they can be asymptotically expanded as the sum of a
series of covariant homogeneous functions of the space-time variables. This
essential result will be extensively applied throughout the paper.

\section{The renormalized energy-momentum}

\subsection{The electromagnetic energy-momentum
in the presence of point charges}

In the elementary case of electrostatics, the energy of the
electric field
associated with a non-singular spatially confined charge distribution $\rho (%
{\bf x})$ is given by
\begin{equation}
U=\frac{1}{2}\int d^{3}{\bf x\,E}^{2}({\bf x}),  \label{1.1}
\end{equation}
where
\begin{equation}
{\bf E}({\bf x})=\frac{1}{4\pi }\int d^{3}{\bf x}^{\prime }\rho ({\bf x}%
^{\prime })\frac{{\bf x-x}^{\prime }}{\left| {\bf x-x}^{\prime
}\right| ^{3}}.  \label{1.2}
\end{equation}
When the charge distribution is constituted by a set of $N$
point charges $q_{1}$, $q_{2}$,$\ldots $, $q_{N}$ located at the points $%
{\bf z}_{1}$, ${\bf z}_{2}$,$\ldots $, ${\bf z}_{N}$, Eq.\ (\ref{1.2}%
) takes the form
\[
{\bf E(x)}=\frac{1}{4\pi }\sum_{n=1}^{N}q_{n}\frac{{\bf
x-z}_{n}}{\left| {\bf x-z}_{n}\right| ^{3}}.
\]
It is well known that in such a case the integral on the
right-hand side of Eq.\ (\ref{1.1}) becomes divergent. However,
introducing a cut-off at a distance $r$ from the point charges,
one can define
\[
\overline{U}(r)\equiv \frac{1}{2}\int d^{3}{\bf x\,E}^{2}({\bf x})
\prod_{n=1}^{N}%
\theta \left( \left| {\bf x-z}_{n}\right| -r\right) ,
\]
where $\theta $ is the usual step function. Then an elementary
integration provides
\begin{equation}
\overline{U}(r)=\frac{1}{8\pi r}\sum_{n=1}^{N}q_{n}^{2}
+\frac{1}{4\pi }\sum_{n<m}\frac{%
q_{n}q_{m}}{\left| {\bf z}_{n}-{\bf z}_{m}\right| }+O(r)\text{\quad for }%
r\rightarrow 0.  \label{1.3}
\end{equation}
It can be seen therefore that the divergent part is only
represented by an irrelevant constant term which can be safely
dropped, leaving us with the familiar Coulomb potential energy:
\begin{equation}
U\left( {\bf z}_{1},\ldots ,{\bf z}_{N}\right) =\frac{1}{4\pi }\sum_{n<m}%
\frac{q_{n}q_{m}}{\left| {\bf z}_{n}-{\bf z}_{m}\right| }.
\label{1.4}
\end{equation}

The above considerations illustrate the obvious fact that the
singularities associated with point charges are completely
harmless in the case of electrostatics, since the physics can be
completely described by finite quantities such as (\ref{1.4}). In
the present paper we shall show the remarkable fact that an
analogous result actually holds also for relativistic
electrodynamics. To this end, it is useful to introduce a general
limiting procedure for functions near a first-order singular
point.

\begin{definition}
Given a function $f(r)$ of a variable $r$, having in a
neighborhood of $r=0$ a behavior of the form
\[
f(r)=\frac{a}{r}+b+O(r),
\]
we define its ``${\cal R}$-limit'' (henceforth indicated by the symbol $%
{\cal R}\lim $) for $r\rightarrow 0$ as
\[
{\cal R}\lim_{r\rightarrow 0}f(r)\equiv \lim_{r\rightarrow 0}\frac{d}{dr}%
\left[ rf(r)\right] =b.
\]
Correspondingly, given a function $G({\bf x})$ of the spatial coordinates $%
{\bf x}$, having in a set $V$ the singular points ${\bf z}_{1}$,
${\bf z}_{2}$,$\ldots $, ${\bf z}_{N}$, and a behavior such that
\begin{equation}
\int_{V} d^{3}{\bf x}\,G({\bf x})\prod_{n=1}^{N}\theta \left( \left| {\bf x-z}%
_{n}\right| -r\right) =\frac{a^{\prime }}{r}+b^{\prime }+O(r),  \label{2.1}
\end{equation}
we define its ``${\cal R}$-integral'' over $V$ as
\[
{\cal R}\int_{V} d^{3}{\bf x}\,G({\bf x})\equiv {\cal R}\lim_{r\rightarrow
0}\int_{V} d^{3}{\bf x}\,G({\bf x})\prod_{n=1}^{N}\theta
\left( \left| {\bf x-z}_{n}\right| -r\right) =b^{\prime }.
\]
\end{definition}

It is clear that the ${\cal R}$-integral defined above differs from
the ordinary integral only in a neighborhood of the singular points.
More precisely, if $V = V_1 \cup V_2$, where
$V_1 \cap V_2 = \O$ and $V_2$ contains no singularities, then
\[
{\cal R}\int_{V} d^{3}{\bf x}\,G({\bf x}) =
{\cal R}\int_{V_1} d^{3}{\bf x}\,G({\bf x}) + \int_{V_2}
d^{3}{\bf x}\,G({\bf x})\,.
\]

Applying the Definition 1 to the electrostatic problem considered at
the beginning, we can write
\[
U\left( {\bf z}_{1},\ldots ,{\bf z}_{N}\right) =\frac{1}{4\pi }\sum_{n<m}%
\frac{q_{n}q_{m}}{\left| {\bf z}_{n}-{\bf z}_{m}\right| }={\cal R}%
\int d^{3}{\bf x}\,(1/2){\bf E}^{2}({\bf x}).
\]
The generalization of the above expression to the electrodynamic case will
be obtained simply by replacing in the integrand the electrostatic energy
density $(1/2){\bf E}^{2}$ with the full electromagnetic energy density $%
T^{00}=(1/2)\left( {\bf E}^{2}+{\bf B}^{2}\right)$. The same procedure can
then be applied also with the momentum density $T^{0k}$. In this way, for a
given inertial reference frame we are led to consider as the electromagnetic
energy-momentum contained at the time $t$ in a volume $V$ the quantity
\begin{equation}
P_{V,{\rm em}}^{\mu }(t)\equiv {\cal R}\int_V d^{3}{\bf x}\,T^{0\mu }({\bf x},t),
\label{2.2}
\end{equation}
where
\begin{equation}
T^{\mu \nu }=F^{\mu \lambda }F_{\,\ \lambda }^{\nu }-\frac{1}{4}g^{\mu \nu
}F^{\eta \lambda }F_{\eta \lambda }  \label{2.3}
\end{equation}
is the familiar electromagnetic energy-momentum tensor (we put $c=1$ and use the
metric $g^{11}=g^{22}=g^{33}=+1$, $g^{00}=-1$). Of course, in order
for the definition (\ref{2.2}) to be meaningful, the integral on the
right-hand side must have a behavior of the form (\ref{2.1}). It is possible
to prove that this requirement is actually satisfied as a consequence
of the Maxwell equations for the electromagnetic field in the presence of
point charges. In fact, the field near a charged particle can be decomposed
as $F^{\mu \nu }=F_{{\rm ext}}^{\mu \nu }+F_{{\rm ret}}^{\mu \nu }$, where $%
F_{{\rm ret}}^{\mu \nu }$ is the retarded field generated by the particle
and $F_{{\rm ext}}^{\mu \nu }$ satisfies the free-field equations in a
neighborhood of the particle. Under the reasonable assumption that $F_{{\rm %
ext}}^{\mu \nu }$ be non-singular in this neighborhood, the
divergent part of
the field can be expressed using the asymptotic expansion for $F_{{\rm adv}%
}^{\mu \nu }$ and $F_{{\rm ret}}^{\mu \nu }$ which is provided in App.\ A.
This can be written as
\begin{equation}
F_{{\rm adv/ret}}^{\mu \nu }(z+x)=(F^{\mu \nu })_{-2}(x)+(F^{\mu \nu
})_{-1}(x)+(F_{{\rm adv/ret}}^{\mu \nu })_{0}(x)+O(x),  \label{2.4}
\end{equation}
with
\begin{equation}
(F)_{-2}(x)=\frac{q}{4\pi y^{3}}u\wedge x,  \label{2.5}
\end{equation}
\begin{equation}
(F)_{-1}(x)=\frac{q}{4\pi y^{3}}\left[ \left( 1-\frac{3x^{2}}{2y^{2}}\right)
(\dot{u}\cdot x)u\wedge x-(u\cdot x)\dot{u}\wedge x
-\frac{x^{2}}{2}u\wedge \dot{u}\right] \,,
\label{2.6}
\end{equation}
while $\left( F_{{\rm adv/ret}}\right) _{0}$, given by Eqs.\ (\ref{A16}) and
(\ref{A17}), is a term that is finite for $x\rightarrow 0$. In the above
formulae, $z$ is the space-time coordinate of the particle at a given proper
time $\tau $ . We have also introduced the four-vectors $u=dz/d\tau $, $%
\dot{u}=du/d\tau $, the scalar $y=\sqrt{x^{2}+(x\cdot u)^{2}}$, and we have used
the notations $u\cdot x=u_{\nu }x^{\nu }$, $x^{2}=x\cdot x=x_{\nu }x^{\nu }$%
, $(u\wedge x)^{\mu \nu }=u^{\mu }x^{\nu }-u^{\nu }x^{\mu }$ and similar.
One can see that, since $y$ is a homogeneous function of $x$ of degree 1,
the term $(F)_{-2}$ is homogeneous of degree $-2$, and moreover it is odd
under the inversion $x\rightarrow -x$ (leaving $u$ and $\dot{u}$ unchanged).
Consequently, when the function, as in Eq.\ (\ref{2.2}), is restricted to
the three-dimensional space with $x^{0}=0$, this term is odd under the space
reflection ${\bf x}\rightarrow -{\bf x}$. Similarly, $(F)_{-1}$ is
homogeneous of degree $-1$ and is even. It follows that $T^{0\mu }$ has a
leading term of order $x^{-4}$ which, when substituted to $G$ in an integral
of the form (\ref{2.1}), gives rise to the term $a^{\prime }/r$ on the
right-hand side. Besides, the term of $T^{0\mu }$\ of order $x^{-3}$ is odd,
so it does not generate any logarithmic term in $r$. One can therefore
conclude that the integral has indeed the required behavior.

It can be observed, using the same arguments, that the value of the
${\cal R}$-integral on the right-hand side of Eq.\ (\ref{2.2})
is unchanged if the cut-off regions near the point charges,
instead of being spherical as in Eq.\ (\ref{2.1}), are chosen with a
different shape, provided that such a shape be invariant
under space inversion with respect to
the particle position. Owing to the form of the expansion (\ref{2.4})--(\ref
{2.6}), it turns out that mathematical manipulations are simpler when
surfaces of the form $y=$ constant are considered. Therefore we shall make
use more often of the equivalent expression
\begin{equation}
P_{V,{\rm em}}^{\mu }(t)={\cal R}\lim_{r\rightarrow 0}
\overline{P}_{V,{\rm em}}^{\mu }(t,r),  \label{2.7}
\end{equation}
with
\begin{equation}
\overline{P}_{V,{\rm em}}^{\mu }(t,r)\equiv \int_V d^{3}{\bf x}\,T^{0\mu }
({\bf x},t)\prod_{n=1}^{N}\theta (y_{n}-r)  \label{2.8}
\end{equation}
and
\begin{equation}
y_{n}\equiv \sqrt{(x-z_{n})^{2}+\left[ u_{n}\cdot (x-z_{n})\right]
^{2}}\,.  \label{2.9}
\end{equation}
Using $z_{n}^{0}=x^{0}=t$, ${\bf z}_{n}={\bf z}_{n}(t)$, ${\bf v}_{n}(t)=d%
{\bf z}_{n}/dt$, $\gamma _{n}=\left( 1-{\bf v}_{n}^{2}\right) ^{-1/2}$, we
can write also
\begin{equation}
y_{n}=\sqrt{({\bf x}-{\bf z}_{n})^{2}+\gamma _{n}^{2}\left[ {\bf v}_{n}\cdot
({\bf x}-{\bf z}_{n})\right] ^{2}},  \label{2.10}
\end{equation}
from which it can be seen that the surfaces $y_n=r$ are just spheres
relativistically contracted by a factor $\gamma$ in the direction of
movement.
As shown in App.\ B, the term of order $1/r$ in
$\overline{P}_{V,{\rm em}}^{\mu }(t,r)$ can be easily
calculated from the expansion (\ref{2.4}). This gives
\begin{eqnarray}
\overline{P}_{V,{\rm em}}^{k}(t,r) &=&\frac{1}{4\pi r}\sum_{{\bf z}_n \in V}
\frac{2}{3}q_{n}^{2} u_{n}^{k}+P_{V,{\rm em}}^k (t)+O(r),  \label{2.12} \\
\overline{P}_{V,{\rm em}}^{0}(t,r) &=&\frac{1}{4\pi r}\sum_{{\bf z}_n \in V}
q_{n}^{2}\left( \frac{2}{3}u_{n}^{0}-\frac{1}{6}\gamma _{n}^{-1}\right)
+P_{V,{\rm em}}^0 (t)+O(r), \label{2.11}
\end{eqnarray}
where the sums on the right-hand sides extend to all the particles
contained in the volume $V$.
One can observe from these expressions that the
divergent terms of the electromagnetic energy-momentum
do {\it not} transform as a four-vector under Lorentz
transformations. In particular, for a single particle with $v \ll 1$,
the $1/r$ terms alone would give rise to the well-known result
$\overline{P}_{V,{\rm em}}^{k}=(4/3)\overline{P}_{V,{\rm em}}^{0}v^k$.
However, these divergent terms depend only on the
particle velocities and contain no information
about the interaction of the particles with each
other or with incident external fields. This suggests that
these terms may be dropped as it was done for the constant divergent term in
Eq.\ (\ref{1.3}). In so doing, we are led directly to the definition (\ref
{2.2}). In the following subsection we are going to provide
a rigorous justification for this definition.

\subsection{The conservation of the renormalized energy-momentum \\
and the Lorentz-Dirac equation}

If we denote with $P^\mu_V (t)$ the total energy-momentum contained
the volume $V$ at the time $t$, in a local theory we must expect that
\begin{equation}
\frac{d}{dt}P^\mu_V (t)=
-\int_{\partial V}d\sigma\, n_i T^{i\mu}({\bf x},t)\,,  \label{conserv}
\end{equation}
where ${\bf n}$ is the outward unit vector normal to the surface
element $d\sigma$. Clearly the right-hand side of the above equation represents
the total flow of energy-momentum through the boundary $\partial V$
of the volume $V$.
We are now going to show how the condition expressed
by Eq.\ (\ref{conserv}),
combined with the definition (\ref {2.2}) for the electromagnetic
energy-momentum, univocally determines the dynamics of the
charged particles in a relativistically covariant way.
However, in order to formulate the equations of motion, we first need to
introduce the fields $\widetilde{F}^{\mu \nu
}\left( z_{n}\right) $, which are defined only on the world-line of the
particles according to the relation
\begin{equation}
\widetilde{F}^{\mu \nu }\left( z_{n}\right)
\equiv {\cal R}\lim_{r\rightarrow 0}
\int_{\left| {\bf x}\right| =r}\frac{d\Omega _{{\bf x}}%
}{4\pi }F^{\mu \nu }\left( {\bf z}_{n}+{\bf x},t\right) ,  \label{2.15}
\end{equation}
with $t=z_{n}^{0}$. The integral in the solid angle on the right-hand side
represents the average of the field
on a spherical surface of radius $r$ centered at ${\bf z}%
_{n}(t)$. Again, one can deduce from Eq.\ (\ref{2.4}) and from symmetry
considerations that the $r$ dependence of the integral is indeed of the form
required for the ${\cal R}$-limit to be properly defined.

\begin{theorem}
If one defines
\begin{equation}
P^\mu_V (t) \equiv P_{V,{\rm em}}^{\mu }(t)+
\sum_{{\bf z}_n \in V}m_n u_n^\mu (t) \,,  \label{PV}
\end{equation}
with $P_{V,{\rm em}}^{\mu }(t)$ given by Eq.\ (\ref{2.2}),
then the condition expressed by Eq.\ (\ref {conserv}) is satisfied
for arbitrary $V$ if and only if the particles obey the equation of motion
\begin{equation}
m_{n}\gamma _{n}\frac{du_{n}^{\mu }}{dt}=q_{n}u_{n,\nu }\widetilde{F}^{\mu
\nu }(z_{n})\,,  \label{2.18}
\end{equation}
with $\widetilde{F}^{\mu \nu }\left( z_{n}\right)$ given by Eq.\
(\ref {2.15}).
\end{theorem}

The total energy-momentum of a system with $N$ charged particles
is obtained from Eq.\ (\ref{PV}) by
taking as the volume $V$ the whole space. This gives
\begin{equation}
P^\mu (t) = P_{{\rm em}}^{\mu }(t)+
\sum_{n=1}^N m_n u_n^\mu (t)\,, \label{2.13}
\end{equation}
with
\begin{equation}
P_{{\rm em}}^{\mu }(t) \equiv {\cal R}\int d^{3}{\bf x}\,T^{0\mu }({\bf x},t)
\,. \label{Pem}
\end{equation}
From Eq.\ (\ref {conserv}) and from the usual assumption that the field
vanishes sufficiently fast at infinity, it follows that $P^\mu (t)$
is a constant of the motion, i.e.\ satisfies
\begin{equation}
\frac{dP^{\mu }}{dt}=0\,.  \label{2.14}
\end{equation}
Furthermore, one can prove that it has the right transformation property
required by special relativity.

\begin{theorem}
The total energy-momentum $P^{\mu }$ given by Eq.\ (\ref {2.13})
transforms as a four-vector under Lorentz transformations.
\end{theorem}

The proofs of Theorems 1 and 2 are presented in Appendixes C and D respectively.
It is important to observe that if we have a single charged particle in
uniform rectilinear motion, and if the field at every point of space reduces
to that produced by the particle (i.e.\ it is equal to the retarded or
advanced fields, which are in this case coincident), then the expression
$\overline{P}_{{\rm em}}^{\mu }(t,r)$, defined by Eq.\ (\ref {2.8})
with $V$ extending to the whole space,
is exactly equal to the $1/r$ term alone.
From the definition of ${\cal R}$-integral it follows that
$P_{{\rm em}}^{\mu }(t)$ vanishes, so that the total energy-momentum is
given by $P^{\mu }=mu^{\mu }$. An analogous result holds for a set of
uniformly moving particles at very large distances from each other. These
considerations imply that the parameters $m_{n}$ appearing in
Eqs.\ (\ref{PV})--(\ref{2.18})
are to be identified with the total physical masses of the particles.

Considering the equation of motion (\ref {2.18}),
we see that the right-hand side has the familiar structure of the Lorentz
force, but with the new finite field $\widetilde{F}^{\mu \nu }$ instead of
the total electromagnetic field $F^{\mu \nu }$, which of course diverges on
the trajectory of the particles. Eq.\ (\ref{2.18}) has the correct
mathematical form that an equation of motion in a theory of interacting
particles and fields is supposed to have. It expresses in fact the
second-order time derivative of the particle's position as a function of the
position, its first-order derivative and the instantaneous field
configuration, since $\widetilde{F}^{\mu \nu }(z_{n})$ is a function of the
field $F^{\mu \nu }({\bf x},t)$ at fixed $t$ in a neighborhood of ${\bf z}%
_{n}(t)$. A complete description of the temporal evolution of the system is
obtained by coupling with Eq.\ (\ref{2.18}) the Maxwell equations
\begin{eqnarray}
\frac{\partial {\bf B}({\bf x},t)}{\partial t}+{\bf \nabla \wedge E}({\bf x}%
,t) &=&0  \label{2.19} \\
\frac{\partial {\bf E}({\bf x},t)}{\partial t}-{\bf \nabla \wedge B}({\bf x}%
,t) &=&-{\bf j}({\bf x},t)=-\sum_{n=1}^{N}q_{n}{\bf v}_{n}(t)\delta \left(
{\bf x}-{\bf z}_{n}(t)\right) ,  \label{2.20}
\end{eqnarray}
with the supplementary equations
\begin{eqnarray}
{\bf \nabla \cdot E}({\bf x},t) &=&\rho ({\bf x},t)=\sum_{n=1}^{N}q_{n}%
\delta \left( {\bf x}-{\bf z}_{n}(t)\right)  \label{2.21} \\
{\bf \nabla \cdot B}({\bf x},t) &=&0  \label{2.22}
\end{eqnarray}
to be regarded as usual as constraints on the initial conditions.

In order to clarify
the meaning of Eq.\ (\ref{2.18}), let us define for a given charged
particle the field
\begin{equation}
\overline{F}^{\mu \nu }\equiv F^{\mu \nu }-\frac{1}{2}\left( F_{\rm %
adv}^{\mu \nu }+F_{\rm ret}^{\mu \nu }\right) ,  \label{2.24}
\end{equation}
which remains finite on the world-line of the particle. Using Eqs.
(\ref{2.4}) and (\ref{A16}) we can then write
\begin{equation}
F^{\mu \nu }(z+x)=\left( F^{\mu \nu }\right) _{-2}(x)+\left(
F^{\mu \nu }\right) _{-1}(x)+\left( F^{\mu \nu }\right) _{0}(x)+%
\overline{F}^{\mu \nu }(z)+O(x)\ ,  \label{2.25}
\end{equation}
where $\left( F^{\mu \nu }\right) _{0}(x)$, given by Eq.\ (\ref{A17}),
is a term that
changes sign under the substitution $x\rightarrow -x$.
Therefore, applying the definition (\ref{2.15}) and taking into account the
spatial antisymmetry of $(F)_{-2}$\ and $(F)_{0}$, we obtain
\begin{equation}
\widetilde{F}^{\mu \nu }(z) =\lim_{r\rightarrow
0}\int_{\left| {\bf x}\right| =r}\frac{d\Omega _{{\bf x}}}{4\pi }\left[
\left( F^{\mu \nu }\right) _{0}(x)+\overline{F}^{\mu \nu }(z)%
\right] =\overline{F}^{\mu \nu }(z)  \label{2.26}
\end{equation}
(a proof of Eq.\ (\ref{2.26}) for the system of reference comoving
with the particle was already given in Ref. \cite{sette}). Since
the definition (\ref{2.24}) is Lorentz covariant, Eq.\
(\ref{2.26}) shows that $\widetilde{F}^{\mu \nu }(z)$
actually transforms like a tensor. Substituting
$\overline{F}^{\mu \nu } $\ to $\widetilde{F}^{\mu \nu }$\ on
the right-hand side of Eq.\ (\ref {2.18}), one can recognize in
the latter a well-known form of the relativistic Lorentz-Dirac
equation. We note also that Eqs.\ (\ref{2.4}) and (\ref{A16}) are
in agreement with the relation first proved by Dirac in \cite{uno}
\[
\frac{1}{2}%
\left[ F_{\rm ret}^{\mu \nu }(x)-F_{\rm adv}^{\mu \nu }(x)\right]
\rightarrow \frac{2}{3}\frac{q}{4\pi }%
\left( u^{\mu }\ddot{u}^{\nu }-u^{\nu }\ddot{u}^{\mu }\right)
\text{ for }x\rightarrow z\ ,
\]
from which it follows that, after decomposing the field as
$F^{\mu \nu }=F_{\rm ext}^{\mu \nu }+F_{\rm ret}^{\mu \nu }$, the
Eq.\ (\ref{2.18}) can be rewritten as
\begin{equation}
m\dot{u}^{\mu }=qu_{\nu }F_{{\rm ext}}^{\mu \nu }
+\frac{2}{3}\frac{q^{2}}{4\pi }(%
\ddot{u}^{\mu }-\dot{u}_{\nu }\dot{u}^{\nu }u^{\mu }),  \label{2.27}
\end{equation}
where use has been made of the relations
$u^{2}=-1$, $0=d(u^{2})/d\tau =2\dot{u}\cdot u$, $0=d(\dot{u}\cdot u)/d\tau =
\dot{u}^{2}+\ddot{u}\cdot u$.
In the familiar form (\ref{2.27}) the Lorentz-Dirac equation
appears explicitly as a {\it third-order} differential equation allowing to
calculate the trajectory of a point charge interacting with an assigned
``external'' field $F_{{\rm ext}}^{\mu \nu }$ (given in general by an
incident radiation field plus the sum of the retarded fields generated by
other charges). On the contrary, in our formulation of the theory Eq.\ (\ref
{2.18}) is a {\it second-order} equation which alone cannot determine the
particle's trajectory, since it is intrinsically coupled with the Eqs.\ (\ref
{2.19}) and (\ref{2.20}) for the total field $F^{\mu \nu }$. The three
equations need be solved together in order to determine the temporal
evolution of the system of interacting particles and fields. One may observe
incidentally that these equations are manifestly invariant under time
reversal, which is not the case for the Lorentz-Dirac equation written in
the form (\ref{2.27}).

The solution of the system of coupled equations (\ref{2.18}) and
(\ref{2.19})--(\ref{2.20})
depends on a set of initial conditions that is entirely specified by the
values, at a given time $\bar{t}$, of ${\bf z}_{n}(\bar{t})$, ${\bf v}_{n}(\bar{t%
})$\ and $F^{\mu \nu }({\bf x},\bar{t})$\ for every ${\bf x}$. These
initial conditions are subject to some relevant constraints,
in addition to those already expressed by Eqs.\ (\ref{2.21}) and (\ref{2.22}%
). The reasonable requirement that $\overline{F}^{\mu \nu }(x)$\ be
non-singular at $z_{n}$ implies that the divergent part of $F^{\mu \nu }(%
{\bf x},\bar{t})$ must have the form given by Eqs.\ (\ref{2.4})--(\ref{2.6}%
). Therefore the term of order $\left| {\bf x}-{\bf z}_{n}(\bar{t})\right|
^{-2}$ is completely determined by ${\bf v}_{n}(\bar{t})$. Furthermore, the
term of order $\left| {\bf x}-{\bf z}_{n}(\bar{t})\right| ^{-1}$ is related
to ${\bf v}_{n}(\bar{t})$ and to the\ zeroth order term $\widetilde{F}^{\mu
\nu }\left( {\bf z}_{n}(\bar{t})\right) $, since the vector $\dot{u}_{n}$
appearing in Eq.\ (\ref{2.6}) must coincide with that resulting from
Eq.\ (\ref{2.18}).

Another subtler constraint on the initial conditions is related to
the well-known problem of the ``run-away'' solutions of the
Lorentz-Dirac equation \cite{uno,tre,treb}. We remark incidentally that
such solutions do not violate the conservation law given by Eq.\
(\ref{2.14}). In fact the unlimited increase towards $+\infty $ of
the particle's kinetic energy is compensated by a corresponding
decrease towards $-\infty $ of the electromagnetic energy $P_{{\rm
em}}^{0}$ (which is not a definite positive quantity, according to
our definition), and analogous statements are valid for the three
spatial components of the momentum. These solutions, which are
therefore completely legitimate from a mathematical standpoint,
are generally dismissed for phenomenological reasons, since
run-away electrons have never been observed.

We remind that, in
the usual treatment of the Lorentz-Dirac equation as a third-order
differential equation, the initial particle acceleration has to be
specified among the initial conditions, together with the
velocity, the position and the incident field. One may then take
advantage of this additional free parameter in order to get rid of
the run-away solutions. For instance, considering a scattering
problem between a particle and a wave-packet initially very far apart,
it is possible in general to obtain a finite result for the
asymptotic momentum of the particle in the final state by
introducing a suitable ``preacceleration'', i.e.\ an acceleration
of the particle that decreases as $\exp (t/\gamma \tau _{0})$ for
$t\rightarrow -\infty $, with $\tau _{0}=q^{2}/6\pi m c^3$
($\simeq 0.62\times 10^{-23}$ sec in the case of an electron), and
that becomes significant starting from a time of order $\tau _{0}$
before the actual interaction between particle and
wave-packet takes place.

In our formulation of
the Cauchy problem for the equations (\ref{2.18}) and
(\ref{2.19})--(\ref{2.20}), since the particle equation of motion
is now of second order, the initial acceleration cannot be
independently assigned, but is determined by the initial value of
the total electromagnetic field according to Eq.\ (\ref{2.18}).
Therefore the requirement that the asymptotic particle momenta be
finite both for $t\rightarrow -\infty $ and for $t\rightarrow
+\infty $ is equivalent to a constraint on the initial values of
the field. The fact that, in the traditional ``third-order''
treatment, the preacceleration is exponentially vanishing for
times $t$ very far in the past with respect to the actual
interaction, within the new ``second-order'' picture translates
into the fact that an equivalently small ``adjustment'' of an
arbitrarily assigned field configuration at the same times $t$ is
generally required in order to satisfy the asymptotic conditions.

It is finally worth mentioning that the run-away behavior cannot
be revealed by a perturbative expansion in the
coupling constant. Therefore the existence of run-away solutions
cannot be claimed to be a formal
drawback of the classical theory of
point charges as compared to perturbative quantum
electrodynamics. It has also been conjectured that a
particular kind of run-away solutions may exist in relativistic
spin 1/2 QED \cite{nove}.

\section{The action principle}

\subsection{The renormalized lagrangian}

A fundamental property of the system of differential
equations constituted by Eqs.\ (\ref{2.18}) and
(\ref{2.19})--(\ref{2.22}) is that it can be
derived from an action principle. A lagrangian formalism can in fact be
introduced in a natural way by taking as configuration variables of the
system the position ${\bf z}_{n}(t)$ of the particles and the four-vector
electromagnetic potential $A^{\mu }({\bf x},t)$ satisfying the relation
\begin{equation}
F^{\mu \nu }=\partial ^{\mu }A^{\nu }-\partial ^{\nu }A^{\mu }.  \label{3.1}
\end{equation}
Given an inertial system of coordinates, we shall define as the total
lagrangian of the system at the time $t$ the functional
\begin{equation}
L=L_{{\rm em}}+L_{{\rm part}}+L_{{\rm int}}\,,  \label{3.2}
\end{equation}
with
\begin{equation}
L_{{\rm em}}=-\frac{1}{4}{\cal R}\int d^{3}{\bf x}\,F_{\mu \nu }({\bf x}%
,t)F^{\mu \nu }({\bf x},t),  \label{3.3}
\end{equation}
\begin{equation}
L_{{\rm part}}=-\sum_{n=1}^{N}\frac{m_{n}}{\gamma _{n}},
\end{equation}
\begin{equation}
L_{{\rm int}}=\sum_{n=1}^{N}\frac{q_{n}}{\gamma _{n}}u_{n}^{\nu }\widetilde{A%
}_{\nu }\left( z_{n}\right) .  \label{3.5}
\end{equation}
The vector $\widetilde{A}^{\nu }\left( z_{n}\right) $ in Eq.\ (\ref{3.5}) is
defined in perfect analogy with Eq.\ (\ref{2.15}):
\begin{equation}
\widetilde{A}^{\nu }\left( z_{n}\right) \equiv {\cal R}\lim_{r\rightarrow
0}\int_{\left| {\bf x}\right| =r}\frac{d\Omega _{{\bf x}}}{4\pi }A^{\nu
}\left( {\bf z}_{n}+{\bf x},t\right) .  \label{3.6}
\end{equation}
One can see therefore that $L$ is indeed a functional of the configuration
variables and of their first time derivatives, taken at a given time $t$. In
general this functional is defined for every field configuration having a
behavior near the singular points such that the ${\cal R}$-integral in Eq.\ (%
\ref{3.3}) and the ${\cal R}$-limit in Eq.\ (\ref{3.6}) are properly defined.

As usual, Eqs.\ (\ref{2.19}) and (\ref{2.22}) are automatically
satisfied as a result of Eq.\ (\ref{3.1}), so that the equations of motion
for the variables ${\bf z}_{n}$ and $A^{\mu }(x)$ reduce to Eqs.\ (\ref{2.18}%
), (\ref{2.20}) and (\ref{2.21}). We are now going to show that these
equations follow from Hamilton's variational principle.

\begin{theorem}
If one defines the action $I$ as
\[
I=\int_{t_{1}}^{t_{2}}L\ dt,
\]
with $L$ given by Eqs.\ (\ref{3.2})--(\ref{3.5}), then the equations of
motion of the system in the time interval $t_{1}<t<t_{2}$ are equivalent to
the condition that $\delta I=0$ for every infinitesimal variation $\delta
A^{\mu }(x)$ and $\delta {\bf z}_{n}(t)$ vanishing at $t=t_{1}$ and $t=t_{2}$%
, and such that the varied trajectories $\left\{ {\bf z}_{n}+\delta {\bf z}%
_{n},A^{\mu }+\delta A^{\mu }\right\} $ remain in the definition
domain of $I$.
\end{theorem}

In order to prove this theorem, we shall
introduce the non-renormalized lagrangian
\begin{equation}
\overline{L}(r)\equiv \overline{L}_{{\rm em}}(r)+L_{{\rm part}}+\overline{L}%
_{{\rm int}}(r),  \label{3.8}
\end{equation}
with
\begin{equation}
\overline{L}_{{\rm em}}(r)=-\frac{1}{4}\int d^{3}{\bf x}\,F_{\mu \nu }({\bf x%
},t)F^{\mu \nu }({\bf x},t)\prod_{n=1}^{N}\theta \left( \left| {\bf x-z}%
_{n}\right| -r\right)  \label{3.9}
\end{equation}
and
\begin{equation}
\overline{L}_{{\rm int}}(r)=\sum_{n=1}^{N}\frac{q_{n}}{\gamma _{n}}%
u_{n}^{\nu }\int_{\left| {\bf x}\right| =r}\frac{d\Omega _{{\bf x}}}{4\pi }%
A_{\nu }\left( {\bf z}_{n}+{\bf x},t\right) .  \label{3.10}
\end{equation}
Since $L={\cal R}\lim_{r\rightarrow 0}\overline{L}(r)$, we can write
\begin{equation}
\overline{L}(r)=(L)_{-1}r^{-1}+L+O(r)\ ,  \label{3.11}
\end{equation}
where the coefficient $(L)_{-1}$ is some other functional
of the configuration variables and of their first time derivatives.
It follows from Eq.\ (\ref{3.11}) that
\begin{equation}
\delta\overline{L}(r)=\delta(L)_{-1}r^{-1}+
\delta L+O(r)\ .  \label{3.11diff}
\end{equation}
Therefore the variations of the renormalized lagrangian $L$ will be
evaluated by calculating the corresponding variations of
$\overline{L}$, and then taking their ${\cal R}$-limit for
$r\rightarrow 0$.

\subsection{Derivation of the field equations from the action principle}

Let us fix at a first stage $\delta {\bf z}_{n} \equiv 0$ and consider only
variations of the electromagnetic potential, which will be represented by
arbitrary infinitesimal test functions $\delta A^{\mu }(x)$ satisfying the
boundary conditions. We want to show that, in order to have identically $%
\delta I=0$, the fields $F^{\mu \nu }$ must define distributions that
satisfy the differential equations (\ref{2.20}) and (\ref{2.21}).
For $\delta {\bf z}_{n}\equiv 0$ we have
\begin{equation}
\delta \overline{L}(r)=-\int d^{3}{\bf x}\,F^{\mu \nu }\partial _{\mu
}\delta A_{\nu }+\sum_{n=1}^{N}\frac{q_{n}}{\gamma _{n}}u_{n}^{\nu }\delta
A_{\nu }(z_{n})+O(r),  \label{3.13}
\end{equation}
where we have used the fact that, assuming $F^{\mu \nu }$ to have at most a
singularity of order $r^{-2}$ in $z_{n}$, the integral on the right-hand
side is convergent, and moreover
\[
\int_{\left| {\bf x}\right| =r}\frac{d\Omega _{{\bf x}}}{4\pi }\delta A_{\nu
}\left( {\bf z}_{n}+{\bf x},t\right) =\delta A_{\nu }(z_{n})+O(r^{2}).
\]
Since on the right-hand side of Eq.\ (\ref{3.13}) there is no term of order
$r^{-1}$, a comparison with Eq.\ (\ref{3.11diff})
shows that $\delta (L)_{-1}=0$ and
\begin{equation}
\delta I=\int_{t_{1}}^{t_{2}}dt\,\delta L=
\int_{t_{1}}^{t_{2}}dt\left[-\int d^{3}{\bf x}\,F^{\mu \nu }\partial
_{\mu }\delta A_{\nu }+\sum_{n=1}^{N}\frac{q_{n}}{%
\gamma _{n}}u_{n}^{\nu }\delta A_{\nu }(z_{n})\right]\,.  \label{3.15}
\end{equation}
Since $\delta A_{\nu }$ are test functions vanishing for $t=t_1$ and
$t=t_2$, Eq.\ (\ref{3.15}) can be rewritten as
\[
\delta I=\int_{t_{1}}^{t_{2}}dt\int d^{3}{\bf x}\,\left[ \partial _{\mu
}F^{\mu \nu }+\sum_{n=1}^{N}\frac{q_{n}}{\gamma _{n}}u_{n}^{\nu }\delta (%
{\bf x}-{\bf z}_{n})\right] \delta A_{\nu }\,,
\]
where the partial derivatives of $F^{\mu \nu }$ are
defined according to the distribution theory. Therefore the
condition $\delta I=0$ for arbitrary $\delta A_{\nu }$ implies
\begin{equation}
\partial _{\mu }F^{\mu \nu }=-\sum_{n=1}^{N}\frac{q_{n}}{\gamma _{n}}%
u_{n}^{\nu }\delta ({\bf x}-{\bf z}_{n})=-j^{\nu },  \label{3.17}
\end{equation}
which is equivalent to Eqs.\ (\ref{2.20}) and (\ref{2.21}). Of course,
assuming that the fields $F^{\mu \nu }$ are represented by ordinary
differentiable functions of $x$ everywhere outside the world-lines of the
particles, Eq.\ (\ref{3.17}) implies in particular that in the same region
such functions satisfy the differential equation $\partial _{\mu }F^{\mu \nu
}=0$ in the usual sense.

It is well-known that the general solution of Eq.\ (\ref{3.17}), when
the particle trajectories are assigned,
is given by the sum of the Li\'{e}nard-Wiechert
fields and a solution of the free-field equations. Using this fact, we
are now going to show that the lagrangian $L$ is indeed correctly defined on
the solutions of the field equations and on the neighboring trajectories
obtained via non-singular variations $\delta A_{\nu }$ of the potential.
This fact ensures the self-consistency of the whole procedure.
First of all, since asymptotic conditions of the type (\ref{2.4}) must be
satisfied by the fields near the particles, we can repeat the same
considerations we made about the ${\cal R}$-integral in Eq.\ (\ref{2.2}),
showing that also the right-hand side of Eq.\ (\ref{3.3}) is correctly
defined. Moreover, using the expansion for the advanced and
retarded vector potentials derived in App.\ A, one can verify that
also the potential $\widetilde{A}_{\nu }\left( z_{n}\right) $
appearing in Eq.\ (\ref{3.5}) is well defined according to Eq.\ (\ref{3.6}).
In particular, defining for a given particle
\begin{equation}
\overline{A}^{\mu }\equiv A^{\mu }-\frac{1}{2}\left(
A_{\rm adv}^{\mu }+A_{\rm ret}^{\mu }\right) ,  \label{3.19}
\end{equation}
one gets from Eqs.\ (\ref{A7}) and (\ref{A9})
\begin{equation}
A^{\mu }(z+x)=\left( A^{\mu }\right) _{-1}(x)+\left( A^{\mu}
\right) _{0}(x)+\overline{A}^{\mu }(z)+O(x),  \label{3.20}
\end{equation}
and from the antisymmetry in $x$ of $\left( A^{\mu }\right) _{0}(x)$ it
follows that
\begin{equation}
\widetilde{A}^{\mu }(z) =\lim_{r\rightarrow 0}\int_{\left|
{\bf x}\right| =r}\frac{d\Omega _{{\bf x}}}{4\pi }\left[ \left( A^{\mu}
\right) _{0}(x)+\overline{A}^{\mu }(z)\right] =
\overline{A}^{\mu }(z),  \label{3.21}
\end{equation}
in close analogy with Eqs.\ (\ref{2.24})--(\ref{2.26}). We remark also that
Eqs.\ (\ref{3.19}) and (\ref{3.21}) ensure that $\widetilde{A}^{\mu }(z)$
behaves as a four-vector under Lorentz transformations.

\subsection{Derivation of the particle equation of motion from the action
principle}

When considering also variations $\delta {\bf z}_{n}(t)\neq 0$,
one has to pay attention to the fact that the quantity
$\widetilde{A}_{\nu }\left( z_{n}\right) $ is defined only on the
world-line of a particle, so that its partial derivatives with
respect to ${\bf z}_{n}$ are not defined in the
usual sense. This is equivalent to the obvious statement that, whenever $%
\delta {\bf z}_{n}(t)\neq 0$ and $\delta A^{\mu }(x)=0$, the varied
trajectories no longer belong to the definition domain of the functional $L$%
. In order to remain within this domain, each variation $\delta {\bf z}%
_{n}(t)\neq 0$ must be accompanied by a simultaneous variation $\delta
A^{\mu }$\ of the potentials, characterized by an appropriate singular
behavior near the particles. For this reason, it is convenient to proceed in
two steps. First, given the boundary conditions at $t=t_{1}$ and $t=t_{2}$,
for each possible trajectory $z_{n}$ of the particles one determines the
potential $A_{z_{n}}^{\mu }(x)$ that minimizes the action $I$ with respect
to the variations of the potential alone: as we have seen in the
previous subsection, $A_{z_{n}}^{\mu }(x)$
is just the potential that solves the Maxwell equations
for those particular particle trajectories. Then one can define the
functional $I^{\prime }\left[ z_{n}\right] $ of the trajectory of the
particles alone, as
\begin{equation}
I^{\prime }\left[ z_{n}\right] \equiv I\left[ z_{n},A_{z_{n}}^{\mu }(x)%
\right] \,.  \label{3.22}
\end{equation}
It is now possible to show that the condition $\delta I^{\prime }=0$, for
arbitrary variations $\delta {\bf z}_{n}(t)$ satisfying $\delta {\bf z}%
_{n}(t_{1})=\delta {\bf z}_{n}(t_{2})=0$, is equivalent to the fact that $%
{\bf z}_{n}(t)$, together with $A_{z_{n}}^{\mu }(x)$, are solutions of Eq.\ (%
\ref{2.18}). Of course, the minima of $I$ must coincide with the minima of $%
I^{\prime }$: therefore the trajectory of particles and fields $\left\{
z_{n},A_{z_{n}}^{\mu }(x)\right\} $ satisfying Eq.\ (\ref{2.18}) is indeed
the trajectory that minimizes $I$ within its whole definition domain,
subject to the assigned boundary conditions.

Keeping these ideas in mind, we shall still write for simplicity $%
A^{\mu }$ instead of $A_{z_n}^{\mu }$, although we shall always
implicitly assume that the field satisfies Eq.\ (\ref{3.17}).
We shall therefore apply the previous formulae
describing the asymptotic behavior of the field near the point
charge. We shall make use also of an expansion for the derivatives
of the potentials that, similarly to Eq.\ (\ref{2.25}), follows
from the results of App.\ A:
\begin{equation}
\partial _{\mu }A_{\nu }(z+x)=\left( \partial _{\mu }A_{\nu }\right)
_{-2}(x)+\left( \partial _{\mu }A_{\nu }\right) _{-1}(x)+\left(
\partial _{\mu }A_{\nu }\right) _{0}(x)+\partial _{\mu
}\widetilde{A}_{\nu }(z)+O(x), \label{3.33}
\end{equation}
where $\left( \partial _{\mu }A_{\nu }\right) _{j}(x)$ is a
homogeneous function of degree $j$ and even (resp. odd) parity for
$j$ odd (resp. even), while
\begin{equation}
\partial _{\mu }\widetilde{A}_{\nu }(z)\equiv {\cal R}\lim_{r\rightarrow
0}\int_{\left| {\bf x}\right| =r}\frac{d\Omega _{{\bf x}}}{4\pi
}\,\partial _{\mu }A_{\nu }\left( {\bf z}+{\bf x},t\right) .
\label{3.34}
\end{equation}
Note that the above equation is to be considered as a definition
since, as we already observed, the partial derivatives of
$\widetilde{A}(z)$ do not exist in the usual sense. However one
can easily show that, in close analogy with Eqs.\ (\ref{2.26}) and
(\ref{3.21}),
\begin{equation}
\partial _{\mu }\widetilde{A}_{\nu }(z)=\partial _{\mu }\overline{A}_{\nu
}(z),  \label{3.34b}
\end{equation}
where the right-hand side represents the usual partial derivative
of the function $\overline{A}(x)$ defined by Eq.\ (\ref{3.19}).

With a rather tedious but straightforward calculation, which is
reported in App. E, one obtains
\begin{eqnarray}
\delta L &=&-\frac{d}{dx^{0}}\left( {\cal R}\int d^{3}{\bf x}\,F^{0j}\delta
A_{j}\right) \nonumber \\
&&+\sum_{n=1}^N \left[
q_n\widetilde{A}_{i}(z_n)\delta v^{i}_n+q_n\gamma^{-1}_n u^{\nu}_n
\partial _{i}\widetilde{A}_{\nu }(z_n)\delta z^{i}_n+m_i u_{n,i}
\delta v^{i}_n \right] \nonumber \\
&=&\frac{d}{dt}\left[ \sum_{n=1}^N
\left( m_n u_{n,i}+q_n\widetilde{A}_{i}(z_n)\right) \delta
z^{i}_n-{\cal R}\int d^{3}{\bf x}\,F^{0j}\delta A_{j}
\right] \nonumber \\
&&-\sum_{n=1}^N
\left[ \frac{d}{dt}\left( m_n u_{n,i}
+q_n\widetilde{A}_{i}(z_n)\right) -q_n\gamma^{-1}_n u^{\nu }_n
\partial _{i}\widetilde{A}_{\nu }(z_n)\right] \delta z^{i}_n\,.
\label{deltal}
\end{eqnarray}
Since $\delta z^i_n=0$ and $\delta A_\nu =0$ for $t=t_1$ and
$t=t_2$, the boundary terms
do not contribute to the variation of the action, and we obtain
\begin{equation}
\delta I =-\int_{t_{1}}^{t_{2}}dt\,\sum_{n=1}^N
\left[ \frac{d}{dt}\left( m_n u_{n,i}
+q_n\widetilde{A}_{i}(z_n)\right) -q_n\gamma^{-1}_n u^{\nu }_n
\partial _{i}\widetilde{A}_{\nu }(z_n)\right] \delta z^{i}_n\,.
\end{equation}

Since it follows from Eqs.\ (\ref{3.21}) and (\ref {3.34b}) that
\[
\frac{d}{dt}\widetilde{A}_{\mu }(z_n)=\frac{d}{dt}\overline{A}_{\mu
}(z_n)=\gamma ^{-1}_n u^{\nu }_n\partial _{\nu }
\overline{A}_{\mu }(z_n)=\gamma
^{-1}_n u^{\nu }_n\partial _{\nu }\widetilde{A}_{\mu }(z_n)\,,
\]
the condition $\delta I=0$ for arbitrary $\delta z^{i}_n$ implies
\begin{equation}
m_n\dot{u}_{n,i}=m_n\gamma_n \frac{du_{n,i}}{dt}=q_n\left(
\partial _{i}\widetilde{A}_{\nu}(z_n)-\partial _{\nu }
\widetilde{A}_{i}(z_n)\right) u^{\nu }_n=q_n\widetilde{F}_{i\nu
}(z_n)u^{\nu }_n\ ,  \label{3.39}
\end{equation}
that coincides with Eq.\ (\ref{2.18}). We have therefore proved that
the dynamics of the system is entirely determined by Hamilton's variational
principle applied to the lagrangian defined by
Eqs.\ (\ref{3.2})--(\ref{3.5}), as stated by Theorem 3.
Of course, it is also possible to deduce directly the conservation laws
of energy and momentum from the translational invariance in space and time
of $L$. This is illustrated in detail in App.\ F.

\subsection{\label{mass1}A comparison with the idea of mass renormalization}

The renormalization of our lagrangian was achieved simply through
the introduction of a ${\cal R}$-integral for the electromagnetic
part in Eq.\ (\ref{3.3}) and a ${\cal R}$-limit of the
electromagnetic potential for the interaction part in Eq.\
(\ref{3.5}). Clearly this procedure has in principle nothing to do
with the renormalization of the particle's mass, which is typical
of other approaches to the argument, and in particular of
standard renormalization techniques of quantum field theories.
Therefore no infinitely negative ``bare mass'' has to be
introduced in our theory, and the difficulty of simultaneously
compensating with such a mass term the non-covariant divergent
parts in Eqs.\ (\ref{2.12})--(\ref{2.11}) is automatically avoided.
It is however interesting to note that the divergent part of Eq.\
(\ref{3.9}) can be easily calculated using Eq.\ (\ref {B9}):
\begin{eqnarray*}
(L_{{\rm em}}) _{-1}r^{-1} &=&-\frac{1}{4}\int d^{3}{\bf x}\,\left( F^{\mu
\nu }\right) _{-2}\left( F_{\mu \nu }\right) _{-2}\theta (y-r) \\
&=&-\frac{1}{4\gamma }\int_{r}^{\infty }y^{2}dy\int d\Omega _{{\bf y}}\left[
-\frac{2q^{2}}{(4\pi )^{2}y^{4}}\right] =\frac{q^{2}}{8\pi \gamma r}
\end{eqnarray*}
(we consider here for simplicity a single charged particle),
so that we have
\begin{equation}
L_{{\rm em}}=\lim_{r\rightarrow 0}\left[ -\frac{1}{4}\int d^{3}{\bf x}%
\,F^{\mu \nu }F_{\mu \nu }\theta (y-r)-\frac{q^{2}}{8\pi \gamma r}\right] \,.
\label{3.40}
\end{equation}
Similarly we have from Eq.\ (\ref{A7})
\[
\int_{y=r}d\Omega _{{\bf y}}\left( A_{\nu }\right) _{-1}=\int_{y=r}d\Omega _{%
{\bf y}}\frac{qu_{\nu }}{4\pi y}=\frac{qu_{\nu }}{r},
\]
so that
\begin{equation}
L_{{\rm int}}=\frac{q}{\gamma }u^{\nu }{\cal R}\lim_{r\rightarrow
0}\int_{y=r}\frac{d\Omega _{{\bf y}}}{4\pi }A_{\nu }({\bf z}+{\bf x}%
)=\lim_{r\rightarrow 0}\left[ \frac{q}{\gamma }u^{\nu }\int_{y=r}\frac{%
d\Omega _{{\bf y}}}{4\pi }A_{\nu }({\bf z}+{\bf x})+\frac{q^{2}}{4\pi
r\gamma }\right] \,.  \label{3.41}
\end{equation}
Therefore, combining Eqs.\ (\ref{3.2}), (\ref{3.40}) and (\ref{3.41}), we
can write
\begin{equation}
L=\lim_{r\rightarrow 0}\left[ -\frac{1}{4}\int d^{3}{\bf x}\,F^{\mu \nu
}F_{\mu \nu }\theta (y-r)+\frac{q}{\gamma }u^{\nu }\int_{y=r}\frac{d\Omega _{%
{\bf y}}}{4\pi }A_{\nu }({\bf z}+{\bf x})-\frac{m_{0}}{\gamma }\right] \,,
\label{3.42}
\end{equation}
with
\[
m_{0}=m-\frac{q^{2}}{8\pi r}.
\]

Clearly, one could interpret Eq.\ (\ref{3.42}) as the lagrangian of a
particle having indeed a divergent negative bare mass $m_{0}$ and carrying a
total electric charge $q$ that, for an observer with the same instantaneous
velocity of the particle, looks uniformly distributed over an infinitesimal
spherical surface of radius $r$. We must however underline that Eq.\ (\ref
{3.42}) is verified only when the fields are already solutions of Maxwell
equations. In general, taking Eq.\ (\ref{3.42}) as a definition would not be
equivalent to the definition of $L$ that we have given in Eqs.\ (\ref{3.2}%
)--(\ref{3.5}). This becomes particularly evident when introducing
the hamiltonian formalism. In fact, since each of the three terms
on the right-hand side of Eq.\ (\ref{3.42}) is divergent, it is
immediate to see that also the conjugate momenta of the particles
$p_{i}=\partial L/\partial v^{i}$ are divergent in the
$r\rightarrow 0$ limit. On the contrary, the lagrangian
(\ref{3.2}) is the sum of three {\it individually finite} terms.
As we shall show in the next section, from this fact it follows
that all the momenta are finite, and therefore an exact
hamiltonian can be defined for the theory.

\section{Euler-Lagrange and Hamilton equations}

It is useful to define the partial derivatives of the lagrangian $L$ with
respect to its arguments, in such a way that the equations of motion can be
formally derived as the usual Euler-Lagrange equations associated with $L$. We
have seen in the previous section that,
in order to remain within the original
definition domain (that we shall call $D_{0}$) of $L$, each variation $%
\delta z_n$ must be accompanied by a variation $\delta A$ of the
electromagnetic potential with an appropriate singular behavior near the
particles. Therefore, defining a partial derivative of the type $\partial
L/\partial z^{i}_n$ for constant potential $A^{\mu }(x)$ requires introducing
an extension of $L$ outside $D_0$. We have to note
first of all that, inside $D_{0}$, the particles'
positions $z_n$ are univocally
determined by the field variables $A^{\mu }(x)$, since the potential is
assumed to be singular just in $z_n$. In other words, as soon as $A^{\mu }(%
{\bf x},t)$ is assigned for every ${\bf x}$, also its singular points, that
will be now called ${\bf w}_n(t)$ to distinguish them in general from the
particles' positions ${\bf z}_n(t)$, are of course determined,
and inside $D_{0}$
it is always assumed that ${\bf z}_n={\bf w}_n$. It is clear now that, in order
to treat ${\bf z}_n$ and $A^{\mu }(x)$ as really independent variables, we
have to define $L$ also for ${\bf z}_n\neq {\bf w}_n$. Having this in mind, the
results of the previous section indicate how to extend in an obvious way the
definition of $L$. We shall put
\begin{equation}
\frac{\partial L_{{\rm em}}}{\partial z^{i}_n}=0,  \label{4.1}
\end{equation}
which is consistent with considering $L_{{\rm em}}$\ as the purely
electromagnetic part of the lagrangian, and
\begin{equation}
\frac{\partial L_{{\rm int}}}{\partial z^{i}_n}=
q_n\frac{u^{\nu }_n}{\gamma_n }%
\partial _{i}\widetilde{A}_{\nu }(z_n),  \label{4.2}
\end{equation}
which is equivalent to extending in a neighborhood of ${\bf w}_n$ the function
$\widetilde{A}_{\nu }(z_n)$ appearing in Eq.\ (\ref{3.5}) according to
\[
\widetilde{A}_{\nu }({\bf z}_n,t)=\widetilde{A}_{\nu }({\bf w}_n,t)
+\partial _{i}\widetilde{A}_{\nu }({\bf w}_n,t)(z^{i}_n-w^{i}_n)
+O\left[ ({\bf z}_n-{\bf w}_n)^{2}\right]
\,.
\]
On the right-hand side of the above equation the
quantities $\widetilde{A}_{\nu }({\bf w}_n,t)$\ and
$\partial _{i}\widetilde{A}_{\nu }({\bf w}_n,t)$\ are
defined according to Eqs.\ (\ref{3.6}) and (\ref{3.34}) respectively,
whereas terms of order $({\bf z}_n-{\bf w}_n)^{2}$\ can be neglected
for our purposes,
since they clearly do not affect the equations of motion.
Besides, the derivatives with
respect to $v^{i}_n=dz^{i}_n/dt$ can be determined unambiguously as
\begin{eqnarray*}
\frac{\partial L_{{\rm em}}}{\partial v^{i}_n} &=&0, \\
\frac{\partial L_{{\rm int}}}{\partial v^{i}_n} &=&q\widetilde{A}_{i}(z_n).
\end{eqnarray*}
We have then, as a result of Eq.\ (\ref{3.39}),
\[
\frac{\partial L}{\partial z^{i}_n}=q_n\frac{u^\nu_n}{\gamma_n }
\partial _{i}\widetilde{A}_{\nu }(z_n)=
\frac{d}{dt}\left[ m_n u_{n,i}+q_n\widetilde{A}_{i}(z_n)%
\right] =\frac{d}{dt}\frac{\partial L}{\partial v^{i}_n}.
\]
Similarly, using Eq.\ (\ref{3.17}), we have
\[
\frac{\delta L}{\delta A_{\mu }(x)}=\partial _{j}F^{j\mu }(x)+
\sum_{n=1}^N q_n\frac{u^{\mu }_n}{\gamma_n }\delta ({\bf x-z}_n)
=-\partial _{0}F^{0\mu }(x)=\frac{d}{dt}\frac{%
\delta L}{\delta \partial _{0}A_{\mu }(x)}\,,
\]
where the functional derivatives of $L$ with respect to $A_{\mu }(x)$ and $%
\partial _{0}A_{\mu }(x)$ are unambiguously defined as distributions in the
usual way. Looking at the first and last members of the two above equations,
the formal structure of Euler-Lagrange equations is directly recognizable.

The hamiltonian functional can be introduced via a Legendre transformation
on $L$, as in ordinary analytical mechanics. As usual, after imposing the
Lorentz gauge condition
\[
\partial _{\mu }A^{\mu }(x)=0,
\]
we shall add a term $-(1/2)\left( \partial _{\mu }A^{\mu }\right) ^{2}$ to
the electromagnetic lagrangian density in order to introduce a conjugate
momentum also for the temporal component $A^{0}(x)$ of the potential. We
shall therefore replace Eq.\ (\ref{3.3}) with
\[
L_{{\rm em}}=-{\cal R}\int d^{3}{\bf x}\,\left\{ \frac{1}{4}F_{\mu \nu }(%
{\bf x},t)F^{\mu \nu }({\bf x},t)+\frac{1}{2}\left[ \partial _{\mu }A^{\mu }(%
{\bf x},t)\right] ^{2}\right\} \,.
\]
As a result, the Euler-Lagrange equation for $A^{\mu }$ becomes
\begin{equation}
\partial _{\nu }\partial ^{\nu }A^{\mu }(x)=
-\sum_{n=1}^N q_n\frac{u^{\mu }_n}{\gamma_n }\delta
({\bf x-z}_n)=-j^{\mu }(x).  \label{4.4}
\end{equation}
Then, after introducing the canonical momenta
\begin{eqnarray}
p_{n,i} &\equiv &\frac{\partial L}{\partial v^{i}_n}=
m_n u_{n,i}+q_n\widetilde{A}_{i}(z_n)
\label{4.5} \\
\pi _{i}(x) &\equiv &\frac{\delta L}{\delta \partial _{0}A^{i}(x)}=F_{0i}(x)
\label{4.6} \\
\pi _{0}(x) &\equiv &\frac{\delta L}{\delta \partial _{0}A^{0}(x)}=\partial
_{\mu }A^{\mu }(x),  \label{4.7}
\end{eqnarray}
we can define the hamiltonian $H$ of the system as
\begin{eqnarray}
H &\equiv &\sum_{n=1}^N p_{n,i}v^{i}_n+{\cal R}\int d^{3}{\bf x}\,
\pi _{\mu }\partial_{0}A^{\mu }-L \nonumber \\
&=&{\cal R}\int d^{3}{\bf x}\,\left( \frac{1}{4}F_{ij}F^{ij}+%
\frac{1}{2}\pi _{\mu }\pi ^{\mu }-\pi ^{i}\partial _{i}A^{0}+\pi
^{0}\partial _{i}A^{i}\right)  \nonumber \\
&&+\sum_{n=1}^{N}\left[ q_{n}\widetilde{A}^{0}(z_{n})+\sqrt{\left(
p_{n,i}-q_{n}\widetilde{A}_{i}(z_{n})\right) \left( p_{n}^{i}-q_{n}%
\widetilde{A}^{i}(z_{n})\right) +m_{n}^{2}}\right] \,.  \label{4.19}
\end{eqnarray}
Hamilton's equations are:
\begin{equation}
\frac{dz^{i}_n}{dt}=\frac{\partial H}{\partial p_{n,i}}
=\frac{p^{i}_n-q_n\widetilde{A}^{i}(z_n)}{\sqrt{\left(
p_{n,k}-q_{n}\widetilde{A}_{k}(z_{n})\right) \left( p_{n}^{k}-q_{n}%
\widetilde{A}^{k}(z_{n})\right) +m_{n}^{2}}}\,,  \label{4.9}
\end{equation}
\begin{equation}
\partial _{0}A^{i}(x)=\frac{\delta H}{\delta \pi _{i}(x)}=\pi
^{i}(x)+\partial ^{i}A_{0}(x),
\end{equation}
\begin{equation}
\partial _{0}A^{0}(x)=\frac{\delta H}{\delta \pi _{0}(x)}=\pi
^{0}(x)-\partial _{i}A^{i}(x),  \label{4.11}
\end{equation}
\begin{equation}
\frac{dp^{i}_n}{dt}=-\frac{\partial H}{\partial z_{n,i}}=
\frac{q_n\partial ^{i}\widetilde{A}^{j}(z_n)\left( p_{n,j}
-q_n\widetilde{A}_{j}(z_n)\right) }{\sqrt{\left(
p_{n,k}-q_{n}\widetilde{A}_{k}(z_{n})\right) \left( p_{n}^{k}-q_{n}%
\widetilde{A}^{k}(z_{n})\right) +m_{n}^{2}}}-
q_n\partial ^{i}\widetilde{A}^{0}(z_n),  \label{4.12}
\end{equation}
\begin{eqnarray}
\partial _{0}\pi ^{i}(x)=-\frac{\delta H}{\delta A_{i}(x)}&=&\partial
_{j}F^{ji}(x)+\partial ^{i}\pi ^{0}(x) \nonumber  \\
&&+\sum_{n=1}^N\frac{q_n\left( p^{i}_n-q_n\widetilde{A}%
^{i}(z_n)\right) \delta ({\bf x-z}_n)}{\sqrt{\left(
p_{n,k}-q_{n}\widetilde{A}_{k}(z_{n})\right) \left( p_{n}^{k}-q_{n}%
\widetilde{A}^{k}(z_{n})\right) +m_{n}^{2}}}\,,
\label{4.13}
\end{eqnarray}
\begin{equation}
\partial _{0}\pi ^{0}(x)=-\frac{\delta H}{\delta A_{0}(x)}=\partial _{i}\pi
^{i}(x)+\sum_{n=1}^N q_n\delta ({\bf x-z}_n).  \label{4.14}
\end{equation}
As usual, Eqs.\ (\ref{4.9})--(\ref{4.11}) are just the inversion of Eqs.\ (%
\ref{4.5})--(\ref{4.7}), whereas Eq.\ (\ref{4.12}) is equivalent to Eq.\ (%
\ref{3.39}) and Eqs.\ (\ref{4.13})--(\ref{4.14}) are equivalent to Eq.\ (\ref
{4.4}).

The theory can be expressed by means of fundamental Poisson brackets between
canonical variables having the familiar form
\begin{eqnarray}
\left\{ z_{n}^{i},p_{m}^{j}\right\} &=&
\delta _{nm}\delta ^{ij}  \label{4.20} \\
\left\{ A^{\mu }({\bf x}),\pi ^{\nu }({\bf y})\right\} &=&g^{\mu \nu }\delta
({\bf x-y}),  \label{4.16}
\end{eqnarray}
all the others being equal to zero. After introducing the total momentum
\begin{equation}
P^{i}\equiv -{\cal R}\int d^{3}{\bf x}\,\pi _{\mu }\partial ^{i}A^{\mu
}+\sum_{n=1}^{N}p_{n}^{i}  \label{4.18}
\end{equation}
and putting $P^{0}\equiv H$, it is straightforward to verify that from Eqs.\
(\ref{4.9})--(\ref{4.16}) it follows
\begin{eqnarray*}
\left\{ z^{i}_n,P^{0}\right\} &=&\frac{\partial P^{0}}
{\partial p_{n,i}}=\frac{dz^{i}_n}{dt} \\
\left\{ z^{i}_n,P^{j}\right\} &=&\frac{\partial P^{j}}
{\partial p_{n,i}}=\delta^{ij} \\
\left\{ p^{i}_n,P^{0}\right\} &=&-\frac{\partial P^{0}}
{\partial z_{n,i}}=\frac{dp^{i}_n}{dt} \\
\left\{ p^{i}_n,P^{j}\right\} &=&-\frac{\partial P^{j}}{\partial z_{n,i}}=0 \\
\left\{ A^{\mu }(x),P^{\nu }\right\} &=&\frac{\delta P^{\nu }}{\delta \pi
_{\mu }(x)}=-\partial ^{\nu }A^{\mu }(x) \\
\left\{ \pi ^{\mu }(x),P^{\nu }\right\} &=&-\frac{\delta P^{\nu }}{\delta
A_{\mu }(x)}=-\partial ^{\nu }\pi ^{\mu }(x),
\end{eqnarray*}
where the members of the last two equations are to be interpreted as
distributions. The above equations show that, as expected, $P^{\mu }$ is the
generator of the space-time translations for the system. Of course, on the
solutions of the equations of motion it is equal to the total
energy-momentum introduced in Sec.\ II. This is explicitly shown in App.\ F.

\section{The renormalization procedure in the wave-vector space}

\subsection{The energy-momentum in terms of the Fourier transform \\
of the electromagnetic field}

Let us operate the Fourier decomposition of the electromagnetic potential
according to the formulae
\begin{mathletters}
\label{5.1}
\begin{eqnarray}
A_{\mu }(x) &=&\frac{1}{2(2\pi )^{3}}\int \frac{d^{3}{\bf k}}{k^{0}}\left[
a_{\mu }({\bf k},t)e^{i{\bf k}\cdot {\bf x}}+a_{\mu }^{\ast }({\bf k},t)e^{-i%
{\bf k}\cdot {\bf x}}\right] \\
\partial _{0}A_{\mu }(x) &=&-\frac{i}{2(2\pi )^{3}}\int d^{3}{\bf k}\left[
a_{\mu }({\bf k},t)e^{i{\bf k}\cdot {\bf x}}-a_{\mu }^{\ast }({\bf k},t)e^{-i%
{\bf k}\cdot {\bf x}}\right] \,,
\end{eqnarray}
with $k^{0}\equiv \left| {\bf k}\right| $. As a consequence of Eqs.\ (\ref
{4.16}) and (\ref{5.1}), the Poisson brackets between the
variables $a_{\mu}$ and $a_{\mu }^{\ast }$ are
\end{mathletters}
\begin{mathletters}
\label{5.3}
\begin{equation}
\left\{ a_{\mu }({\bf k}),a_{\nu }({\bf k}^{\prime })\right\} =\{a_{\mu
}^{\ast }({\bf k}),a_{\nu }^{\ast }({\bf k}^{\prime })\}=0
\end{equation}
\begin{equation}
\left\{ a_{\mu }({\bf k}),a_{\nu }^{\ast }({\bf k}^{\prime })\right\}
=-2(2\pi )^{3}ig_{\mu \nu }k^{0}\delta ({\bf k-k}^{\prime }).
\end{equation}
\end{mathletters}

It is well-known that for a free field one can express the
electromagnetic energy-momentum as
\[
P_{{\rm free}}^{\mu }=\frac{1}{2(2\pi )^{3}}\int \frac{d^{3}{\bf k}}{k^{0}}%
k^{\mu }a^{\nu }({\bf k})a_{\nu }^{\ast }({\bf k}).
\]
It is not surprising that, when point-like charged particles are present,
also the above integral becomes divergent.
The same is true for the integrals on the
right-hand side of the Fourier expansions (\ref{5.1}) when ${\bf x}$ takes
the value of the position ${\bf z}_{n}(t)$ of one of the particles.
This can be verified using the
following asymptotic expansion of $a^{\nu }({\bf k},t)$ for $\left| {\bf k}%
\right| \rightarrow \infty $, which is derived in App.\ G:
\begin{equation}
a^{\nu }({\bf k},t)=\sum_{n=1}^{N}\left[ \left( a_{n}^{\nu }\right) _{-1}(%
{\bf k},t)+\left( a_{n}^{\nu }\right) _{0}({\bf k},t)\right] e^{-i{\bf k}%
\cdot {\bf z}_{n}}+O\left( \left| {\bf k}\right| ^{-3}\right) \,,
\label{5.18}
\end{equation}
with
\begin{eqnarray}
\left( a_{n}\right) _{-1} &=&-q_{n}\frac{u_{n}}{k\cdot u_{n}}\,,
\label{5.19} \\
\left( a_{n}\right) _{0} &=&-iq_{n}\frac{\left( k\cdot \dot{u}_{n}\right)
u_{n}-\left( k\cdot u_{n}\right) \dot{u}_{n}}{\left( k\cdot u_{n}\right) ^{3}}\,.
\label{5.20}
\end{eqnarray}
Clearly, Eq.\ (\ref{5.18}) is the direct counterpart of the expansions (\ref
{3.20}) and (\ref{3.33}) of $A^{\nu }(z+x)$ and $\partial _{\mu }A^{\nu
}(z+x)$ for $\left| {\bf x}\right| \rightarrow 0$. We have in particular
\begin{eqnarray}
\left( A_{n}^{\mu }\right) _{j}(x) &=&\frac{1}{2(2\pi )^{3}}\int \frac{d^{3}%
{\bf k}}{k^{0}}\left[ \left( a_{n}^{\mu }\right) _{j}({\bf k},t)e^{i{\bf k}%
\cdot {\bf x}}+\left( a_{n}^{\mu }\right) _{j}^{\ast }({\bf k},t)e^{-i{\bf k}%
\cdot {\bf x}}\right]  \label{5.21} \\
\left( \partial _{\nu }A_{n}^{\mu }\right) _{j-1}(x) &=&\frac{i}{2(2\pi )^{3}%
}\int \frac{d^{3}{\bf k}}{k^{0}}k_{\nu }\left[ \left( a_{n}^{\mu }\right)
_{j}({\bf k},t)e^{i{\bf k}\cdot {\bf x}}-\left( a_{n}^{\mu }\right)
_{j}^{\ast }({\bf k},t)e^{-i{\bf k}\cdot {\bf x}}\right] \,.  \label{5.22}
\end{eqnarray}
It can be noticed that $\left( a_{n}\right) _{j}$ is a homogeneous function
of ${\bf k}$ of degree $-j-2$. Besides, it is real for $j=-1$ and purely
imaginary for $j=0$.

We can however redefine the divergent
integrals in ${\bf k}$ space by means of a renormalization procedure
equivalent to that introduced in Sec.\ II
for the corresponding expressions in spatial coordinates.

\begin{definition}
Given a function $f(\Lambda)$ having for $\Lambda \rightarrow
\infty$ a behavior of the form
\[
f(\Lambda)=c\Lambda + d + O(\Lambda ^{-1})\,,
\]
we define its ${\cal R}$-limit for $\Lambda \rightarrow \infty$ as
\[
{\cal R}\lim_{\Lambda \rightarrow \infty}f(\Lambda)\equiv d\,.
\]
Correspondingly, given a function $g({\bf k})$ having a behavior such that
\[
\int_{\left| {\bf k}\right| <\Lambda }d^{3}{\bf k}\,g({\bf k})=c^\prime\Lambda
+d^\prime+O(\Lambda ^{-1})\text{ \quad for }\Lambda \rightarrow \infty ,
\]
we define its ${\cal R}$-integral over the ${\bf k}$ space as
\begin{equation}
{\cal R}\int d^{3}{\bf k}\,g({\bf k})\equiv {\cal R}\lim_{\Lambda
\rightarrow \infty}\int_{\left| {\bf k}\right| <\Lambda }d^{3}{\bf k}\,
g({\bf k})=d^\prime.  \label{5.12}
\end{equation}
\end{definition}

Using this mathematical tool one can express in terms of the variables $%
a_{\mu }$ and $a_{\mu }^{\ast }$ the canonical energy-momentum four-vector
for the system of interacting particles and field, as stated by the
following theorem.

\begin{theorem}
The four-vectors $P^{\mu }$ and $\widetilde{A}^{\mu }(z_{n})$,
given respectively by Eqs.\ (\ref{4.19}), (\ref{4.18}) and (\ref{3.6}),
can be equivalently expressed by the equations
\begin{eqnarray}
P^{i} &=&\frac{1}{2(2\pi )^{3}}{\cal R}\int \frac{d^{3}{\bf k}}{k^{0}}%
k^{i}a^{\nu }({\bf k},t)a_{\nu }^{\ast }({\bf k},t)+\sum_{n=1}^{N}p_{n}^{i}%
\,,  \label{5.13} \\
P^{0} &=&\frac{1}{2(2\pi )^{3}}{\cal R}\int d^{3}{\bf k}\,a^{\nu }({\bf k}%
,t)a_{\nu }^{\ast }({\bf k},t)  \nonumber \\
&&+\sum_{n=1}^{N}\left[ q_{n}\widetilde{A}^{0}(z_{n})+\sqrt{\left(
p_{n,i}-q_{n}\widetilde{A}_{i}(z_{n})\right) \left( p_{n}^{i}-q_{n}%
\widetilde{A}^{i}(z_{n})\right) +m_{n}^{2}}\right]  \label{5.14}
\end{eqnarray}
and
\begin{equation}
\widetilde{A}_{\mu }(z_{n})=\frac{1}{2(2\pi )^{3}}{\cal R}\int \frac{d^{3}%
{\bf k}}{k^{0}}\left[ a_{\mu }({\bf k},t)e^{i{\bf k}\cdot {\bf z}%
_{n}}+a_{\mu }^{\ast }({\bf k},t)e^{-i{\bf k}\cdot {\bf z}_{n}}\right] \,.
\label{5.15}
\end{equation}
\end{theorem}

The proof of Theorem 4 is given in App. H.
By making use of Eqs.\ (\ref{5.3}), (\ref{5.14}) and (\ref{5.15}), one
readily obtains the equation of motion of the $a^{\mu }$:
\begin{eqnarray*}
\frac{d}{dt}a^{\mu }({\bf k},t) &=&\left\{ a^{\mu }({\bf k},t),P^{0}\right\}
=-ik^{0}a^{\mu }({\bf k},t)+\frac{i}{2(2\pi )^{3}k^{0}}\sum_{n=1}^{N}q_{n}%
\frac{u_{n}^{\mu }}{\gamma _{n}}e^{-i{\bf k}\cdot {\bf z}_{n}} \\
&=&-ik^{0}a^{\mu }({\bf k},t)+\frac{i}{2(2\pi )^{3}k^{0}}\int d^{3}{\bf x}%
\,j^{\mu }({\bf x},t)e^{-i{\bf k}\cdot {\bf x}},
\end{eqnarray*}
which is of course in accordance with Eqs.\ (\ref{4.4}) and (\ref{5.1}).
We remark also that, with exactly the same procedure followed to
derive Eq.\ (\ref{5.15}), one can prove that the quantity $\partial _{\nu }%
\widetilde{A}_{\mu }(z_{n})$ defined by Eq.\ (\ref{3.34}) can be expressed
as
\[
\partial _{\nu }\widetilde{A}_{\mu }(z_{n})=\frac{i}{2(2\pi )^{3}}{\cal R}%
\int \frac{d^{3}{\bf k}}{k^{0}}k_{\nu }\left[ a_{\mu }({\bf k},t)e^{i{\bf k}%
\cdot {\bf z}_{n}}-a_{\mu }^{\ast }({\bf k},t)e^{-i{\bf k}\cdot {\bf z}_{n}}%
\right] \,.
\]

\subsection{Another remark on the renormalization of mass}

We saw in Sec.\ II that the expression for the electromagnetic
energy-momentum in ${\bf x}$ space can be most conveniently handled when the
cut-off regions in the integration domain are taken as ellipsoids like in
Eqs.\ (\ref{2.7}) and (\ref{2.8}).
Similarly, in some cases the ${\cal R}$-integrals
in ${\bf k}$ space appearing in Eqs.\ (\ref{5.13})--(\ref{5.14})
may be more suitably
expressed by choosing a different cut-off instead of the sphere of
radius $\Lambda $ introduced in Eq.\ (\ref{5.12}). A particularly
interesting situation occurs when a single charged particle is present. In
such a case, putting
\[
\overline{a}^{\nu }(k,t)\equiv a^{\nu }({\bf k},t)e^{i{\bf k}\cdot {\bf z}}\,,
\]
with $k^{0}\equiv \left| {\bf k}\right| $, it can be easily seen that $%
\overline{a}^{\nu }$ transforms as a four-vector, as it can already be
argued from the form of the right-hand sides of Eqs.\ (\ref{5.19}) and (\ref
{5.20}). It follows that also the expression
\begin{equation}
\overline{K}_{\rm em}^\mu(t,\Lambda ) \equiv \frac{1}{2(2\pi
)^{3}}\int \frac{d^{3}{\bf k}}{k^{0}}k^{\mu }a^{\nu }({\bf
k},t)a_{\nu }^{\ast } ({\bf k},t)\theta \left( \Lambda +k_{\lambda
}u^{\lambda }(t)\right)  \label{5.39}
\end{equation}
transforms as a four-vector, since it can be rewritten as
\[
\overline{K}_{\rm em}^\mu(t,\Lambda )\equiv \frac{1}{(2\pi
)^{3}}\int d^{4}k\,k^{\mu }\overline{a}^{\nu
}(k,t)\overline{a}_{\nu }^{\ast }(k,t)\delta (k^{2})\theta
(k^{0})\theta \left( \Lambda +k_{\lambda }u^{\lambda
}(t)\right)\,.
\]
We observe that the integration domain in Eq.\ (\ref{5.39})
is characterized by the condition
\[
\left| {\bf k}\right| -{\bf k}\cdot {\bf v}<\frac{\Lambda }{\gamma }\,,
\]
which corresponds to the volume contained into a rotational ellipsoid with a
focus at the origin. Repeating the procedure used in App. H
to obtain Eqs.\ (\ref{5.29}) and (\ref{5.30}),
it is easy to see that
\[
{\cal R}\lim_{\Lambda \rightarrow \infty } \overline{K}_{\rm
em}^\mu (t,\Lambda )=
\frac{1}{2(2\pi )^{3}}{\cal R}\int \frac{d^{3}{\bf k}%
}{k^{0}}k^{\mu }a^{\nu }({\bf k},t)a_{\nu }^{\ast }({\bf k},t)\ .
\]
Furthermore, for ${\bf v}(t)=0$ we get from Eq.\ (\ref{5.19})
\begin{eqnarray*}
&&\frac{1}{2(2\pi )^{3}}\int \frac{d^{3}{\bf k}}{k^{0}}k^{\mu }\left( a^{\nu
}\right) _{-1}\left( a_{\nu }\right) _{-1}^{\ast }\theta \left( \Lambda
+k_{\lambda }u^{\lambda }(t)\right) \\
&=&-\frac{q^{2}}{2(2\pi )^{3}}\int d^{3}{\bf k}\frac{k^{\mu }}{\left(
k^{0}\right) ^{3}}\theta \left( \Lambda -k^{0}\right) =-\frac{q^{2}}{4\pi
^{2}}\Lambda u^{\mu }(t).
\end{eqnarray*}
Thanks to the Lorentz covariance, we can immediately generalize the above
result for arbitrary ${\bf v}(t)$, obtaining
\begin{equation}
\overline{K}_{\rm em}^\mu(t,\Lambda ) = -\frac{q^{2}}{4\pi ^{2}}%
\Lambda u^{\mu }(t)+
\frac{1}{2(2\pi )^{3}}{\cal R}\int \frac{d^{3}{\bf k}%
}{k^{0}}k^{\mu }a^{\nu }({\bf k},t)a_{\nu }^{\ast }({\bf
k},t)+O(\Lambda ^{-1})\ .  \label{5.40}
\end{equation}

A similar procedure can be followed for the ${\cal R}$-integral that
expresses $\widetilde{A}_{\mu }(z)$ according to Eq.\ (\ref{5.15}). Applying
again the covariant cut-off we can define the four-vector
\begin{eqnarray}
A_{\mu }(z,\Lambda ) &\equiv &\frac{1}{2(2\pi )^{3}}\int \frac{d^{3}{\bf k}}{%
k^{0}}\left[ a_{\mu }({\bf k},t)e^{i{\bf k}\cdot {\bf z}}+a_{\mu }^{\ast }(%
{\bf k},t)e^{-i{\bf k}\cdot {\bf z}}\right] \theta \left( \Lambda
+k_{\lambda }u^{\lambda }(t)\right)  \nonumber \\
&=&\frac{1}{(2\pi )^{3}}\int d^{4}k\,\left[ \overline{a}_{\mu }(k,t)
+\overline{a}_{\mu }^{\ast }(k,t) \right]
\delta (k^{2})\theta (k^{0})\theta \left( \Lambda +k_{\lambda
}u^{\lambda }(t)\right) .  \label{5.41}
\end{eqnarray}
Then, proceeding as in Eqs.\ (\ref{5.36})--(\ref{5.38}), it is immediate to
see that
\[
\widetilde{A}_{\mu }(z)={\cal R}\lim_{\Lambda \rightarrow \infty }A_{\mu
}(z,\Lambda ).
\]
Moreover, since for ${\bf v}(t)=0$%
\begin{eqnarray*}
&&\frac{1}{2(2\pi )^{3}}\int \frac{d^{3}{\bf k}}{k^{0}}\left[ \left( a_{\mu
}\right) _{-1}({\bf k},t)e^{i{\bf k}\cdot {\bf z}}+\left( a_{\mu }\right)
_{-1}^{\ast }({\bf k},t)e^{-i{\bf k}\cdot {\bf z}}\right] \theta \left(
\Lambda +k_{\lambda }u^{\lambda }(t)\right) \\
&=&\frac{q}{(2\pi )^{3}}\int d^{3}{\bf k}\frac{u_{\mu }(t)}{\left(
k^{0}\right) ^{2}}\theta \left( \Lambda -k^{0}\right) =\frac{q}{2\pi ^{2}}%
\Lambda u_{\mu }(t),
\end{eqnarray*}
the Lorentz covariance implies that for arbitrary ${\bf v}(t)$%
\begin{equation}
A_{\mu }(z,\Lambda ) = \frac{q}{2\pi ^{2}}\Lambda u_{\mu }(t)+%
\widetilde{A}_{\mu }(z)+O(\Lambda ^{-1})\text{ \quad for }\Lambda
\rightarrow \infty .  \label{5.42}
\end{equation}

Putting together Eqs.\ (\ref{5.13}), (\ref{5.14}), (\ref{5.40})
and (\ref {5.42}), we can finally write
\begin{eqnarray*}
P^{\mu } &=&mu^{\mu }(t)+q\widetilde{A}^{\mu }(z)+
\frac{1}{2(2\pi )^{3}}{\cal R}\int \frac{d^{3}{\bf k}%
}{k^{0}}k^{\mu }a^{\nu }({\bf k},t)a_{\nu }^{\ast }({\bf k},t)
\\
&=&\lim_{\Lambda \rightarrow \infty }\left[ m_{0}u^{\mu }(t)+qA^{\mu
}(z,\Lambda )+\frac{1}{2(2\pi )^{3}}\int \frac{d^{3}{\bf k}}{k^{0}}k^{\mu
}a^{\nu }({\bf k},t)a_{\nu }^{\ast }({\bf k},t)\theta \left( \Lambda
+k_{\lambda }u^{\lambda }(t)\right) \right] \,,
\end{eqnarray*}
with $m_{0}=m-\left( q^{2}/4\pi ^{2}\right) \Lambda $. We obtain
therefore a result very similar to that discussed in Sec.
\ref{mass1}: when a proper cut-off is chosen and only fields that
already obey Maxwell equations are considered, the method of the
${\cal R}$-integrals leads to expressions that formally correspond
to the introduction of an infinite renormalization of the
particle's mass. The analogy with the situation occurring with
covariant regularization methods in quantum field theory is
evident. It will be analyzed more closely in the following
section.

\section{The comparison with quantum electrodynamics}

In the previous sections of this paper we have provided a complete
diverge-free formulation of the classical electrodynamics of point charges.
In the light of this result it is of course extremely interesting to
reconsider the problem of the relationship between classical and quantum
theories. It is in fact well known that the difficulties related to
ultraviolet divergences in quantum electrodynamics have been up to now
circumvented only in the framework of a perturbation expansion in the
coupling constant $\alpha =e^{2}/4\pi \hbar c\simeq 1/137$. Finite
non-perturbative expressions for the equations of motion or the hamiltonian
operator have not yet been established, and it is even doubtful that they
actually exist within the boundaries of pure electrodynamics. In connection
with this problem, the implications of our new way of formulating the
classical theory can be directly investigated by making use of the
correspondence principle. Of course, in the quantum theory one encounters
also the class of divergences associated with vacuum polarization, that have
no direct equivalent in the classical theory. We shall not here take them
into consideration, and we shall limit our analysis to the simplest
situation in which the analogy with the classical case can be significant,
i.e.\ the electron self-energy to the lowest order in $\alpha $. We shall
assume that the charged particles are represented by spin $1/2$ fermions
(electrons), and therefore described as the quanta of a Dirac field $\psi $.
We shall follow the scheme of canonical quantization and, by analogy with
the classical Poisson brackets (\ref{4.16}) and (\ref{4.20}), we shall
postulate the usual fundamental equal-time commutation or anticommutation
relations between field operators.

Considering the usual Gell-Mann and Low perturbative expansion of the
two-point fermionic Green function $iS_{F}^{\prime }(x)=\langle 0|T\left[
\psi (x)\overline{\psi }(0)\right] |0\rangle $, the contribution of order $%
\alpha $ (using natural units with $c=\hbar =1$) can be expressed as
\begin{equation}
iS_{F}^{(2)}(x)=\frac{(-i)^{2}}{2}\int_{-\infty }^{+\infty
}dt_{1}\int_{-\infty }^{+\infty }dt_{2}\,\langle 0|T\left[
H_{I}(t_{1})H_{I}(t_{2})\psi (x)\overline{\psi }(0)\right] |0\rangle _{(0)}\
,  \label{6.1}
\end{equation}
where the subscript (0) indicates that the matrix element involves the
non-interacting vacuum state and field operators, and only connected Feynman
diagrams have to be taken into account. By analogy with Eq.\ (\ref{4.19}) or
(\ref{5.14}), the interaction hamiltonian is obtained by substituting the
field $\widetilde{A}^{\mu }$ in place of $A^{\mu }$\ into the familiar
expression of the non-renormalized theory. This gives
\begin{equation}
H_{I}(t)=-\int d^{3}{\bf x}\,\widetilde{A}_{\mu }(x)j^{\mu }(x)=-ie\int d^{3}%
{\bf x}\,\widetilde{A}_{\mu }:\overline{\psi }\gamma ^{\mu }\psi :\,,
\label{6.2}
\end{equation}
where $\widetilde{A}^{\mu }$ is given by a ${\cal R}$-integral of the form (%
\ref{5.15}). The natural way to generalize our notion of ${\cal R}$-integral
when the integrand is an operator in a Hilbert space is to perform the
limiting procedure in the weak sense, i.e.\ on the matrix elements of the
operator between any pair of states. This directly provides the prescription
for calculating $iS_{F}^{(2)}$. One first has to replace the field $%
\widetilde{A}_{\mu }(x)$ in Eq.\ (\ref{6.2}) with the expression
\begin{equation}
A_{\mu }(x,\Lambda )=\frac{1}{2(2\pi )^{3}}\int \frac{d^{3}{\bf k}}{k^{0}}%
\left[ a_{\mu }({\bf k})e^{ik\cdot x}+a_{\mu }^{\ast }({\bf k})e^{-ik\cdot x}%
\right] \theta \left( \Lambda -f({\bf k})\right) ,  \label{6.3}
\end{equation}
where $f({\bf k})$ is an appropriate function of ${\bf k}$. Then, after
taking the expectation value on the non-interacting vacuum state of the
time-ordered product indicated in Eq.\ (\ref{6.1}), one has to evaluate a
properly defined ${\cal R}$-limit of the resulting expression for $\Lambda
\rightarrow \infty $. One may notice the direct correspondence between Eq.\ (%
\ref{6.3}) and Eq.\ (\ref{5.41}) when putting $f({\bf k})=-k_{\lambda
}u^{\lambda }$ and considering that, since the field operators in Eq.\ (\ref
{6.1}) satisfy the free-field equations, one can make the substitution
\[
a_{\mu }({\bf k},t)\rightarrow a_{\mu }({\bf k})e^{-ik^{0}t}.
\]

It is easy to see that as a result of these prescriptions the expectation
value in Eq.\ (\ref{6.1}) has to be calculated by making use of the
following free propagators:
\begin{eqnarray*}
\langle 0|T\left[ A^{\mu }(x,\Lambda )A^{\nu }(0,\Lambda )\right] |0\rangle
_{(0)} &=&ig^{\mu \nu }\int \frac{d^{4}k}{(2\pi )^{4}}e^{ik\cdot
x}D_{F}(k,\Lambda )\equiv ig^{\mu \nu }D_{F}(x,\Lambda ) \\
\langle 0|T\left[ \psi (x)\overline{\psi }(0)\right] |0\rangle _{(0)}
&=&i\int \frac{d^{4}p}{(2\pi )^{4}}e^{ip\cdot x}S_{F}(p)\equiv iS_{F}(x),
\end{eqnarray*}
with
\[
D_{F}(k,\Lambda )\equiv -\frac{1}{k^{2}+\lambda ^{2}-i\varepsilon }\theta
\left( \Lambda -f({\bf k})\right)
\]
and
\[
S_{F}(p)=\frac{ip\cdot \gamma -m}{p^{2}+m^{2}-i\varepsilon }.
\]
As usual, a photon mass term $\lambda $ has been included to remove infrared
divergences, but it is expected that the limit $\lambda \rightarrow 0$ can
ultimately be taken for all physically significant quantities. By applying
Wick's theorem one finds for the Fourier transform of Eq.\ (\ref{6.1}):
\[
S_{F}^{(2)}(p)\equiv \int d^{4}x\,S_{F}^{(2)}(x)e^{-ip\cdot
x}=S_{F}(p)\Sigma ^{(2)}(p)S_{F}(p),
\]
with
\begin{equation}
\Sigma ^{(2)}(p)={\cal R}\lim_{\Lambda \rightarrow \infty }\Sigma
^{(2)}(p,\Lambda )  \label{6.6}
\end{equation}
and
\[
\Sigma ^{(2)}(p,\Lambda )=i\frac{\alpha }{4\pi ^{3}}\int
d^{4}k\,D_{F}(k,\Lambda )\gamma _{\mu }S_{F}(p-k)\gamma ^{\mu }.
\]
In the last equation, the integration in $dk^{0}$ can be performed by
closing the integration path in the complex plane and applying the residue
theorem. This gives
\begin{eqnarray*}
\Sigma ^{(2)}(p,\Lambda ) &=&\frac{\alpha }{2\pi ^{2}}\int d^{3}{\bf k}
\,\theta \left( \Lambda -f({\bf k})\right) \Bigg\{\left[
\frac{1}{k^{0}}\frac{%
i(p-k)\cdot \gamma +2m}{(p-k)^{2}+m^{2}}\right] _{k^{0}=\sqrt{{\bf k}%
^{2}+\lambda ^{2}}} \\
&&+\left[
\frac{1}{k^{0}-p^{0}}\frac{%
i(p-k)\cdot \gamma +2m}{k^{2}+\lambda ^{2}}\right] _{k^{0}=p^{0}+\sqrt{({\bf %
k-p})^{2}+m^{2}}}\Bigg\} \ .
\end{eqnarray*}
It can be easily checked that,
at variance with the expressions (\ref{5.39})
and (\ref{5.41}) for the classical self-energy of a point charge,
in the integrand of the above equation
the leading terms of order $\left| {\bf k}\right| ^{-2}$ cancel each
other, whereas the non-vanishing sum of the terms of order $\left| {\bf k}%
\right| ^{-3}$ gives rise to a contribution of order $\alpha \ln \Lambda $
to the electron self-energy.
Considering the case $p^{2}<0$, we shall initially assume
to choose the reference frame for which ${\bf p}=0$, $p^{0}=\sqrt{-p^{2}}$.
Taking $f({\bf k})=\left| {\bf k}\right| /\beta $, where $\beta $ is an
arbitrary parameter, the integrals can thus be easily evaluated. Then, by
imposing the Lorentz covariance, one can generalize the result for arbitrary
$p$. This provides
\begin{equation}
\Sigma ^{(2)}(p,\Lambda )=\frac{\alpha }{2\pi }\left[ m\left( 3\ln \frac{%
2\beta \Lambda }{m}-\frac{1}{2}\right) +\left( ip\cdot \gamma +m\right)
\left( \ln \frac{2\beta \Lambda }{m}-2\ln \frac{m}{\lambda }+\frac{3}{2}%
\right) \right] +\Sigma ^{c(2)}(p),  \label{6.10}
\end{equation}
where $\Sigma ^{c(2)}(p)$ is expressible as a power series containing second
and higher powers of $\left( ip\cdot \gamma +m\right) $ and, as it can be
easily checked, coincides with the finite one-loop electron self-energy
provided by the usual renormalization theory. The form of Eq.\ (\ref{6.10})
obviously suggests the generalization of the notion of ${\cal R}$-limit
which may be applied in Eq.\ (\ref{6.6}): given a function $F(\Lambda )$
having a behavior for of the type
\[
F(\Lambda )=c\ln \Lambda +d+O(\Lambda ^{-1})\text{ \quad for }\Lambda
\rightarrow \infty ,
\]
we shall define
\[
{\cal R}\lim_{\Lambda \rightarrow \infty }F(\Lambda )=d.
\]
We obtain in this way
\begin{equation}
\Sigma ^{(2)}(p)=A^{(2)}+B^{(2)}\left( ip\cdot \gamma +m\right) +\Sigma
^{c(2)}(p),  \label{6.11}
\end{equation}
with
\begin{eqnarray*}
A^{(2)} &=&\frac{\alpha }{2\pi }m\left( 3\ln \frac{2\beta }{m}-\frac{1}{2}%
\right) \\
B^{(2)} &=&\frac{\alpha }{2\pi }\left( \ln \frac{2\beta }{m}-2\ln \frac{m}{%
\lambda }+\frac{3}{2}\right) \,.
\end{eqnarray*}

One can see from Eq.\ (\ref{6.11}) that $A^{(2)}$ represents a finite
electromagnetic mass of the electron. If $m$ is identified with the physical
mass, the above result amounts to introducing in the hamiltonian a bare mass
$m_{0}=m-A^{(2)}+O\left( \alpha ^{2}\right) $. Besides, $B^{(2)}$ can be
considered as the second-order contribution to a finite (apart from infrared
divergences) fermion wave-function renormalization constant $%
Z_{2}=1-B^{(2)}+O\left( \alpha ^{2}\right) $. One may observe that the
arbitrary finite parameter $\beta $ only appears in the unobservable
quantities $m_{0}$ and $Z_{2}$: therefore it simply allows for a finite
renormalization of the theory, once the ultraviolet divergences have been
removed by the ${\cal R}$-limit procedure. One can then conclude that with
the present method the familiar results of perturbative QED at order $\alpha
$ are actually recovered without the need of introducing divergent
parameters in the hamiltonian.

Of course, in order to really establish on a firm basis the
correspondence between classical and quantum electrodynamics, one
should be able to obtain such a result to any order in
perturbation theory. However, an important difference between the
two theories is given by the fact that in the quantum case the
full electron self-energy contains an infinite number of
contributions of any order in $\alpha $. When a manifestly
covariant cut-off $\Lambda $ is used (as in the Pauli-Villars
regularization scheme) the contribution of order $\alpha ^{n}$
contains powers of $\ln \left( \Lambda /m\right) $ up to $\ln
^{n}\left( \Lambda /m\right) $. Therefore the simple logarithmic
dependence which we have found for the lowest-order contribution
in Eq.\ (\ref{6.10}) cannot be assumed to represent the behavior
of the exact electron self-energy as a function of $\Lambda $. We
have just seen how the form of this behavior (which in turn is
strictly related to the large momentum behavior of the
renormalized self-energy) affects the correct definition of the
${\cal R}$-limits that may appear in the equations of motion and
the hamiltonian operator of QED. These considerations therefore
seem to indicate that the problem of explicitly providing a finite
and non-perturbative mathematical formulation of quantum
electrodynamics is in any case ultimately related to the
investigation of the short-distance behavior of the theory.

\appendix

\section{Asymptotic expansion of the field \\
near a point charge}

We want to construct a covariant asymptotic expansion for the
retarded and advanced electromagnetic potentials near a point
$z(\tau )$ of the world-line of a particle with proper time
$\tau$. Such an expansion will have the form
\begin{equation}
A_{{\rm adv/ret}}^{\mu }(z+x)\sim (A_{{\rm adv/ret}}^{\mu })_{-1}(x)
+(A_{{\rm adv/ret}}^{\mu })_{0}(x)+(A_{{\rm adv/ret}}^{\mu })_{1}(x)
+\ldots \,, \label{A1}
\end{equation}
where each term $(A_{{\rm adv/ret}}^{\mu })_{j}(x)$ is a homogeneous
function of $x$ of degree $j$, and contains the time derivatives
$d^{k}z /d\tau ^{k}$ up to the order $k=j+2$.

Clearly, the following two requirements must be satisfied:
\begin{enumerate}
\item  When fixing $\overline{x}\equiv z+x$, since the left-hand side of
Eq.\ (\ref{A1}) becomes of course independent of $\tau $, the series on the
right-hand side must as a whole have the same property.
\item  For $x^{2}=x_{\nu }x^{\nu }=0$ and $x^{0}<0$ (resp.\ $x^{0}>0$), the
well-known expression of the Li\'{e}nard-Wiechert advanced (resp.\ retarded)
potentials must be obtained:
\begin{eqnarray*}
A_{{\rm adv}}^{\mu }(z+x)&=&\frac{q}{4\pi }\frac{u^{\mu }}{x_{\nu }u^{\nu}}
\text{ \quad for }x^{2}=0,\ x^{0}<0 \\
A_{{\rm ret}}^{\mu }(z+x)&=&-\frac{q}{4\pi }\frac{u^{\mu }}{x_{\nu }u^{\nu}}%
\text{ \quad for }x^{2}=0,\ x^{0}>0.
\end{eqnarray*}
\end{enumerate}

Conversely, it is easy to see that if we manage to obtain a series
satisfying the requirements 1 and 2, then Eq.\ (\ref{A1}) will hold for
arbitrary $x$. Let us introduce the scalar
\begin{equation}
y=\sqrt{x^{2}+(x\cdot u)^{2}}  \label{A3}
\end{equation}
and the four-vector
\begin{equation}
\overline{y}^{\mu }=x^{\mu }+x_{\lambda }u^{\lambda }u^{\mu }.  \label{A4}
\end{equation}
We note immediately that
\begin{equation}
\overline{y}_{\mu }u^{\mu }=0,  \label{A5}
\end{equation}
\begin{equation}
\overline{y}_{\mu }\overline{y}^{\mu }=\overline{y}_{\mu }x^{\mu }=x_{\mu
}x^{\mu }+\left( x_{\nu }u^{\nu }\right) ^{2}=y^{2}.
\end{equation}
If we put
\begin{equation}
\left( A_{{\rm adv}}^{\mu }\right) _{-1}(x)=\left( A_{{\rm ret}}^{\mu
}\right) _{-1}(x)=\frac{q}{4\pi }\frac{u^{\mu }}{y}\equiv (A^{\mu })_{-1}(x),
\label{A7}
\end{equation}
we get a term of degree $-1$ which clearly satisfies the requirement 2.
Furthermore, for fixed $\overline{x}$\ we have
\[
\frac{dx^{\mu }}{d\tau }=\frac{d}{d\tau }\left[ \overline{x}^{\mu }-z^{\mu
}(\tau )\right] =-u^{\mu }(\tau ),
\]
\[
\frac{dy}{d\tau }=\frac{\partial y}{\partial x^{\nu }}\frac{dx^{\nu }}{d\tau
}+\frac{\partial y}{\partial u^{\nu }}\frac{du^{\nu }}{d\tau }=\frac{1}{y}%
\left( -\overline{y}_{\nu }u^{\nu }+x_{\lambda }u^{\lambda }x_{\nu }
\dot{u}^{\nu}\right) =\frac{x\cdot u}{y}x\cdot \dot{u},
\]
where the last equality follows from Eq.\ (\ref{A5}). Therefore:
\begin{equation}
\frac{d}{d\tau }(A^{\mu })_{-1}(x)=\frac{q}{4\pi }\left[
\frac{\dot{u}^{\mu }}{y}%
-(x\cdot u)(x\cdot \dot{u})\frac{u^{\mu }}{y^{3}}\right] \,.  \label{A8}
\end{equation}
In order that the property 1 be satisfied, $(A_{{\rm adv/ret}}^{\mu
})_{0}(x) $ must give rise, when differentiated with respect to $\tau $, to
a term of order $x^{-1}$ that exactly cancels that given by Eq.\ (\ref{A8}).
Furthermore, because of requirement 2, we must have $\left( A_{{\rm adv}%
}^{\mu }\right) _{0}(x)=0$ for $x^{2}=0$, $x^{0}<0$, and $\left( A_{{\rm ret}%
}^{\mu }\right) _{0}(x)=0$ for $x^{2}=0$, $x^{0}>0$. It is straightforward
to verify that these conditions are fulfilled by
\begin{equation}
(A_{{\rm adv/ret}}^{\mu })_{0}(x)=(A^{\mu })_{0}(x)\pm
\frac{q}{4\pi }\dot{u}^{\mu}, \label{A9}
\end{equation}
with
\begin{equation}
\frac{4\pi }{q}(A^{\mu })_{0}(x)=-\frac{u\cdot x}{y}\dot{u}^{\mu }-
\frac{x^{2}}{2y^{3}}(\dot{u}\cdot x)u^{\mu }  \label{A10}
\end{equation}
(in formulae like (\ref{A9}),
the upper and lower signs of the symbols $\pm $ and $\mp $
refer to the advanced and retarded fields respectively).
One can then iterate the procedure in order to obtain the higher order
terms. In particular, the term of degree 1 is found to be
\[
(A_{{\rm adv/ret}}^{\mu })_{1}(x)=(A^{\mu })_{1}(x)\mp \frac{q}{4\pi }\left[
\left( \ddot{u}\cdot x-\dot{u}^{2}u\cdot x\right) \frac{u^{\mu }}{3}+
(\dot{u}\cdot x)\dot{u}^{\mu }+(u\cdot x)\ddot{u}^{\mu }\right] \,,
\]
with
\begin{eqnarray*}
\frac{4\pi }{q}(A^{\mu })_{1}(x) &=&\left[ \frac{3x^{4}}{8y^{5}}(\dot{u}\cdot
x)^{2}+\left( 1+\frac{x^{2}}{2y^{2}}\right) \frac{u\cdot x}{3y}\ddot{u}\cdot
x-\left( 2-\frac{x^{2}}{y^{2}}-\frac{x^{4}}{4y^{4}}\right) \frac{y}{6}
\dot{u}^{2}\right] u^{\mu } \\
&&+\left( 1+\frac{x^{2}}{2y^{2}}\right) \frac{u\cdot x}{y}(\dot{u}\cdot x)
\dot{u}^{\mu}+\left( 1-\frac{x^{2}}{2y^{2}}\right) y\ddot{u}^{\mu }.
\end{eqnarray*}

From Eq.\ (\ref{A1}) one can directly obtain, using Eq.\ (\ref{3.1}), an
analogous expansion for the advanced and retarded fields:
\[
F_{{\rm adv/ret}}^{\mu \nu }(z+x)=(F_{{\rm adv/ret}}^{\mu \nu })_{-2}(x)+(F_{%
{\rm adv/ret}}^{\mu \nu })_{-1}(x)+(F_{{\rm adv/ret}}^{\mu \nu
})_{0}(x)+\ldots \,,
\]
with
\[
(F_{{\rm adv/ret}}^{\mu \nu })_{j-1}(x)=\partial ^{\mu }(A_{{\rm adv/ret}%
}^{\nu })_{j}(x)-\partial ^{\nu }(A_{{\rm adv/ret}}^{\mu })_{j}(x).
\]
This provides
\begin{equation}
(F_{{\rm adv}})_{-2}(x)=(F_{{\rm ret}})_{-2}(x)=\frac{q}{4\pi y^{3}}u\wedge
x\equiv (F)_{-2}(x),  \label{A14}
\end{equation}
\begin{eqnarray}
&&(F_{{\rm adv}})_{-1}(x) =(F_{{\rm ret}})_{-1}(x)  \nonumber \\
&=&\frac{q}{4\pi y^{3}}\left[ \left( 1-\frac{3x^{2}}{2y^{2}}\right)
(\dot{u}\cdot x)u\wedge x-(u\cdot x)\dot{u}\wedge x
-\frac{x^{2}}{2}u\wedge \dot{u}\right] \equiv
(F)_{-1}(x)  \label{A15}
\end{eqnarray}
and
\begin{equation}
(F_{{\rm adv/ret}}^{\mu \nu })_{0}(x)=(F^{\mu \nu })_{0}(x)\mp \frac{2}{3}%
\frac{q}{4\pi }(u^{\mu }\ddot{u}^{\nu }-u^{\nu }\ddot{u}^{\mu }),
\label{A16}
\end{equation}
with
\begin{eqnarray}
\frac{4\pi }{q}(F)_{0} &=&\frac{3x^{2}}{2y^{5}}(\dot{u}\cdot x)\left[
\frac{x^{2}}{%
2}u\wedge \dot{u}+(u\cdot x)\dot{u}\wedge x\right]  \nonumber \\
&&+\frac{x^{2}}{2y^{5}}\left[ 3\left( \frac{5x^{2}}{4y^{2}}%
-1\right) (\dot{u}\cdot x)^{2}+\frac{x^{2}}{4}\dot{u}^{2}
+(u\cdot x)(\ddot{u}\cdot x)\right] u\wedge x  \nonumber \\
&&-\frac{x^{2}}{2y^{3}}\ddot{u}\wedge x+\frac{1}{3y}\left( 2+\frac{x^{2}}{%
y^{2}}\right) (u\cdot x)u\wedge \ddot{u}.  \label{A17}
\end{eqnarray}

\section{Divergent terms of the non-renormalized electromagnetic
energy-momentum}

Let us introduce, on the space at fixed $t$ for the laboratory system,
the spatial coordinates ${\bf y}_{n}$\ corresponding to a
reference system moving with the instantaneous velocity of the $n$-th
particle:
\begin{equation}
{\bf y}_{n}={\bf x-z}_{n}+(\gamma _{n}-1)\frac{({\bf x-z}_{n})\cdot {\bf v}%
_{n}}{{\bf v}_{n}^{2}}{\bf v}_{n}\,.  \label{2.16}
\end{equation}
The modulus of the vector ${\bf y}_{n}$ is equal to the scalar $y_{n}$ given
by Eq.\ (\ref{2.10}), so that we can write $y_{n}=\left| {\bf y}_{n}\right| $.
Inverting Eq.\ (\ref{2.16}), performing the translation ${\bf x}-{\bf z}%
_{n}\rightarrow {\bf x}$ and dropping the index $n$, one gets
\begin{equation}
{\bf x}={\bf y}+\left( \frac{1}{\gamma }-1\right) \frac{{\bf y}\cdot {\bf v}%
}{{\bf v}^{2}}{\bf v},  \label{B1}
\end{equation}
from which it follows
\begin{equation}
{\bf x}\cdot {\bf v}=\frac{1}{\gamma }{\bf y}\cdot {\bf v},
\end{equation}
\begin{equation}
{\bf x}^{2}=y^{2}-({\bf y}\cdot {\bf v})^{2}.  \label{B3}
\end{equation}
Besides, from Eqs.\ (\ref{B1}) and (\ref{A4}) one obtains for $x^{0}=0$
\begin{equation}
\overline{y}^{i}=y^{i}+(\gamma -1)\frac{y_{k}v^{k}}{v^{2}}v^{i}.  \label{B5}
\end{equation}
Defining the solid angle $d\Omega _{{\bf y}}$ such that
$y^{2}d\Omega _{\bf y}dy=d^{3}{\bf y}=\gamma d^{3}{\bf x}$, one
obtains immediately
\begin{equation}
\int_{y=r}\frac{d\Omega _{{\bf y}}}{4\pi }y_{i}y_{j}=\frac{\delta _{ij}}{3}%
r^{2}.  \label{B4}
\end{equation}
Then, from Eqs.\ (\ref{B1}), (\ref{B5}) and (\ref{B4}), it is
straightforward to derive that
\begin{equation}
\int_{y=r}\frac{d\Omega _{{\bf y}}}{4\pi }x_{i}\overline{y}_{j}=\frac{\delta
_{ij}}{3}r^{2}.  \label{B6}
\end{equation}

Let us now apply the above formulae to evaluate the $1/r$ term of the
expression (\ref{2.8}). From Eqs.\ (\ref{2.3}) and (\ref{2.5}) it follows
that the leading term of $T^{0k}$ can be expressed as
\[
(T^{0k})_{-4}=(F_{\,\ j}^{0})_{-2}(F^{kj})_{-2}=\frac{q^{2}}{4\pi y^{6}}%
\gamma x_{j}(u^{k}\overline{y}^{j}-u^{j}\overline{y}^{k}),
\]
where we have used the fact that $u\wedge x=u\wedge \overline{y}$. By using
Eq.\ (\ref{B6}) it is easy to obtain that the corresponding leading term of $%
\overline{P}_{{\rm em}}^{\mu }(t,r)$ is given by
\begin{equation}
\int d^{3}{\bf x}(T^{0k})_{-4}\theta (y-r)=\gamma ^{-1}\int_{r}^{\infty
}y^{2}dy\int d\Omega _{{\bf y}}(T^{0k})_{-4}=\frac{q^{2}}{4\pi r}\frac{2}{3}%
u^{k},  \label{B8}
\end{equation}
in accordance with Eq.\ (\ref{2.12}). We have also
\begin{eqnarray}
\left( F^{\mu \nu }\right) _{-2}\left( F_{\mu \nu }\right) _{-2} &=&\frac{%
2q^{2}}{(4\pi )^{2}y^{6}}u_{\mu }\overline{y}_{\nu }(u^{\mu }\overline{y}%
^{\nu }-u^{\nu }\overline{y}^{\mu })=-\frac{2q^{2}}{(4\pi )^{2}y^{4}}\,,
\label{B9} \\
(F^{0i})_{-2}(F_{\,\ i}^{0})_{-2} &=&\frac{q^{2}}{(4\pi )^{2}y^{6}}\gamma
^{2}{\bf x}^{2},
\end{eqnarray}
so that, using Eqs.\ (\ref{2.3}), (\ref{B3}) and (\ref{B4}),
\begin{eqnarray}
\int d^{3}{\bf x}(T^{00})_{-4}\theta (y-r) &=&\gamma ^{-1}\int_{r}^{\infty
}y^{2}dy\int d\Omega _{{\bf y}}(T^{00})_{-4}  \nonumber \\
&=&\frac{q^{2}}{4\pi }\int_{r}^{\infty }\frac{dy}{y^{4}}\int \frac{d\Omega _{%
{\bf y}}}{4\pi }\left( \gamma y^{2}-\gamma y_{i}v^{i}y_{j}v^{j}-\frac{y^{2}}{%
2\gamma }\right)  \nonumber \\
&=&\frac{q^{2}}{4\pi r}\left( \gamma -\frac{\gamma }{3}{\bf v}^{2}-\frac{1}{%
2\gamma }\right) =\frac{q^{2}}{4\pi r}\left( \frac{2}{3}\gamma -\frac{1}{%
6\gamma }\right) \,,  \label{B11}
\end{eqnarray}
to be compared with Eq.\ (\ref{2.11}).

\section{Proof of Theorem 1}

Let us consider a volume $V$ containing a single charged particle.
Omitting the particle's index we can write
\begin{eqnarray}
\frac{d}{dt}\overline{P}_{V,{\rm em}}^{\mu}(t,r) &=&
\frac{d}{dt}\int_V d^{3}{\bf x}%
\,T^{\mu 0}\theta (y-r)=\int_V d^{3}{\bf x}\,\partial _{0}T^{\mu 0}\theta (y-r)
+\int_V d^{3}{\bf x}\,T^{\mu 0}\partial _{0}\theta (y-r)  \nonumber \\
&=&-\int_V d^{3}{\bf x}\,\partial _{i}T^{\mu i}\theta (y-r)+\int_V d^{3}{\bf x}%
\,T^{\mu 0}\partial _{0}\theta (y-r)  \nonumber \\
&=& -\Phi^\mu_V (t) +
\int_V d^{3}{\bf x}\,T^{\mu \nu}\partial _{\nu }\theta (y-r) 
= -\Phi^\mu_V (t) + \int_V d^{3}{\bf x}%
\,T^{\mu \nu}\partial _{\nu }y\,\delta (y-r)\,,  \label{C1}
\end{eqnarray}
where
\[
\Phi^\mu_V (t)\equiv \int_{\partial V}d\sigma\, n_i T^{i\mu}({\bf x},t)
\]
and we have used the equation $\partial _{\nu }T^{\nu \mu }(x)=0$,
which as a consequence of Maxwell equations holds for
every $x$ outside the particle's world-line.
From Eq.\ (\ref{2.10}) one obtains
\begin{equation}
\partial _{i}y\equiv \frac{\partial y}{\partial x^{i}}=\frac{1}{y}\left[
x_{i}-z_{i}+(x_{j}-z_{j})u^{j}u_{i}\right] \,,  \label{C2}
\end{equation}
\begin{equation}
\partial _{0}y\equiv \left. \frac{\partial y}{\partial t}\right| _{{\bf x}}=%
\frac{\partial y}{\partial z^{j}}\frac{dz^{j}}{dt}+\frac{\partial y}{%
\partial u^{j}}\frac{du^{j}}{dt}=-\frac{\partial y}{\partial x^{j}}%
u^{j}\gamma ^{-1}+\frac{1}{y}(x_{k}-z_{k})u^{k}(x_{j}-z_{j})
\dot{u}^{j}\gamma ^{-1}.  \label{C3}
\end{equation}
We can now perform the translation ${\bf x-z}\rightarrow {\bf x}$ and use
the same notation of App.\ A. Then, applying the definition (\ref{A4}) with
$x^{0}=0$, we can rewrite Eqs.\ (\ref{C2}) and (\ref{C3}) as
\begin{equation}
\partial _{i}y=\frac{\overline{y}_{i}}{y}\,, \label{C2bis}
\end{equation}
\begin{equation}
\partial _{0}y=\frac{1}{\gamma y}\left( -\overline{y}%
_{j}u^{j}+x_{j}u^{j}x_{h}\dot{u}^{h}\right) . \label{C3bis}
\end{equation}
Therefore, making in the last term of Eq.\ (\ref{C1}) the change of
integration variables (\ref{B1}) we obtain
\begin{eqnarray}
\frac{d}{dt}\overline{P}_{V,{\rm em}}^{\mu}(t,r)+\Phi^\mu_V (t)&=&
\gamma ^{-1}\int_{0}^{\infty }y^{2}dy\int
d\Omega _{{\bf y}}\,T^{\mu \nu}\partial _{\nu }y\,\delta (y-r) \nonumber \\
&=& r\gamma ^{-1}\int_{y=r}d\Omega
_{{\bf y}}\left( T^{\mu i}\overline{y}_{i}-T^{\mu 0}\overline{y}_{j}u^{j}
\gamma^{-1}+T^{\mu 0}x_{j}u^{j}x_{h}\dot{u}^{h}\gamma ^{-1}\right)
\,.  \label{C5}
\end{eqnarray}
In order to evaluate the behavior of the above expression for $r\rightarrow
0 $, it is useful to expand the field near the particle as in Eq.\ (\ref
{2.25}):
\[
F(x)=(F)_{-2}(x)+(F)_{-1}(x)+(F)_{0}(x)+\widetilde{F}+O(x),
\]
where $(F)_{-2}$, $(F)_{-1}$ and $(F)_{0}$ are given respectively by Eqs.\ (%
\ref{A14}), (\ref{A15}) and (\ref{A17}), while according to Eq.\ (\ref{2.26}%
) $\widetilde{F}\equiv \widetilde{F}(0)=\overline{F}(0)$. Using Eq.\ (\ref
{2.3}) one can perform a similar expansion of the integrand on the
right-hand side of Eq.\ (\ref{C5}). Then, since $(F)_{-2}$ and $(F)_{0}$ are odd
under space inversion, while $(F)_{-1}$ is even, it is easy to see that many
terms of this expansion are spatially odd, and therefore give no
contribution to the integral. Retaining only the terms with even parity, we
obtain
\begin{equation}
\frac{d}{dt}\overline{P}_{V,{\rm em}}^{\mu}(t,r)+\Phi^\mu_V (t)
=(W^{\mu})_{-1}(t)r^{-1} +(W^{\mu})_{0}(t)+O(r),  \label{C7}
\end{equation}
with
\begin{equation}
(W^{\mu})_{-1}(t)r^{-1}=r\gamma ^{-1}\int_{y=r}d\Omega _{{\bf
y}}\left[ (T^{\mu i})_{-3}\overline{y}_{i}-(T^{\mu
0})_{-3}\overline{y}_{j}u^{j}\gamma ^{-1}+(T^{\mu
0})_{-4}x_{j}u^{j}x_{h}\dot{u}^{h}\gamma ^{-1}
\right] \,,  \label{C8}
\end{equation}
\begin{eqnarray}
(W^{\mu})_{0}(t) = r\gamma ^{-1}\int_{y=r}d\Omega _{{\bf
y}}&\Bigg\{&
\widetilde{F}^{\mu\nu }(F_{\,\ \nu }^{i})_{-2}+(F^{\mu\nu })_{-2}\widetilde{F}%
_{\,\ \nu }^{i}+\frac{g^{\mu i}}{2}\widetilde{F}^{\lambda \nu }
(F_{\lambda \nu })_{-2}  \nonumber \\
&&-\left[\widetilde{F}^{\mu\nu} (F_{\,\ \nu }^{0})_{-2}
+(F^{\mu\nu })_{-2}\widetilde{F}_{\,\ \nu }^{0}+
\frac{g^{\mu 0}}{2}\widetilde{F}^{\lambda \nu }(F_{\lambda \nu })_{-2}
\right]\gamma ^{-1}u^{i}\Bigg\} \overline{y}_{i}\,.  \label{C9}
\end{eqnarray}
In Eq.\ (\ref{C8}), $\left( T^{\mu \nu }\right) _{-4}$ and $\left( T^{\mu
\nu }\right) _{-3}$ are respectively the homogeneous terms of degree $-4$
and $-3$ of the energy-momentum tensor, which can be simply expressed as
combinations of $(F)_{-2}$ and $(F)_{-1}$. However, by comparing
Eq.\ (\ref{C7}) with Eqs.\ (\ref{2.12})--(\ref{2.11}) we obtain directly
\begin{eqnarray}
(W^{k})_{-1}r^{-1}&=&\frac{q^{2}}{4\pi r}\frac{2}{3}
\dot{u}^{k}\gamma ^{-1}, \label{C10} \\
(W^{0})_{-1}r^{-1}&=&\frac{q^{2}}{4\pi r}
\left[\frac{2}{3}\dot{u}^{0}\gamma ^{-1}-\frac{1}{6}\frac{d(\gamma
^{-1})}{dt}
\right]\,, \label{C10bis} \\
(W^{\mu})_{0}&=&\frac{d}{dt}P_{V,{\rm em}}^{\mu} +\Phi^\mu_V \,.
\label{C11}
\end{eqnarray}
Of course, Eqs.\ (\ref{C10})--(\ref{C10bis})
can also be verified starting from Eq.\ (\ref{C8})
and carrying out a straightforward although somewhat lengthy calculation,
which may only serve as a consistency check for the whole procedure. The
integrations on the right-hand side of Eq.\ (\ref{C9}) can instead be done
immediately using Eqs.\ (\ref{2.5}) and (\ref{B6}). Taking into account Eq.\
(\ref{C11}) this provides
\begin{equation}
\frac{d}{dt}P_{V,{\rm em}}^{\mu}
+\Phi^\mu_V = -q\widetilde{F}^{\mu\nu }u_{\nu }\gamma ^{-1}.
\label{C12}
\end{equation}
The equation of motion for the charged particle can then be obtained by
postulating that, in accordance with Eqs.\ (\ref{conserv}) and (\ref{PV}),
\[
\frac{d}{dt}\left( P_{V,{\rm em}}^{\mu}+mu^{\mu}\right) =
-\Phi^\mu_V \,.
\]
Using Eq.\ (\ref{C12}) this yields
\[
m\dot{u}^{\mu }=q\widetilde{F}^{\mu \nu }u_{\nu }\,,
\]
which corresponds to Eq.\ (\ref{2.18}).

Vice versa, let us consider a
volume $V$ containing an arbitrary number of particles, all obeying the
equation of motion (\ref{2.18}). Then,
by decomposing $V$ as the union of a due number of subsets each
containing a single particle, and then applying to each of them the equation
(\ref{C12}), it is immediate to prove that Eq.\ (\ref{conserv}) is satisfied.

\section{Proof of Theorem 2}

\subsection{The one-particle case}

Let us consider a tube $T$ of arbitrary shape in four-dimensional
space-time, containing the world-line of a single particle of charge $q$.
Let us suppose also that a given point $P(\overline{\tau })$ of this
world-line, corresponding to a value $\overline{\tau }$ of the proper time,
is assigned the coordinates $z^{\mu }=z^{\prime \mu }=0$ in two inertial
reference frames $O$ and $O^{\prime }$. We shall define $\pi $ and $\pi
^{\prime }$ as the three-dimensional spaces characterized by the equations $%
x^{0}=0$ and $x^{\prime 0}=0$ respectively. Clearly both spaces intersect
the world-line at the point $P(\overline{\tau })$. Let us also define the
vector
\[
J^{\mu }(x)\equiv T^{\mu \nu }(x)b_{\nu }\,,
\]
where $b$ is an arbitrary constant four-vector. From the local conservation
of the electromagnetic energy-momentum tensor $T^{\mu \nu }$, it follows
immediately that $\partial _{\mu }J^{\mu }=0$ everywhere in the tube $T$
except on the particle's world-line. For $r$ much smaller than the diameter
of the tube $T$, we have via a by-parts integration
\begin{eqnarray}
0 &=&\int_{T}d^{4}x\,\partial _{\mu }J^{\mu }\theta (x^{0})\theta
(-x^{\prime 0})\theta (y-r)  \nonumber \\
&=&-\int_{T\cap \pi }d^{3}{\bf x}\,J^{0}\theta (-x^{\prime 0})\theta
(y-r)+\int_{T\cap \pi ^{\prime }}d^{3}{\bf x}^{\prime }\,J^{\prime 0}\theta
(x^{0})\theta (y-r)  \nonumber \\
&&+\int_{\partial T}d\sigma \,J^{\mu }n_{\mu }\theta (x^{0})\theta
(-x^{\prime 0})-\int_{T}d^{4}x\frac{J^{\mu }\overline{y}_{\mu }}{y}\theta
(x^{0})\theta (-x^{\prime 0})\delta (y-r),  \label{D1}
\end{eqnarray}
where $n$ is the unitary four-vector normal to the element $d\sigma $ of the
tube's surface $\partial T$, $y$ and $\overline{y}^{\mu }$ are again defined
according to Eqs.\ (\ref{A3})--(\ref{A4}), and we have used the relation $%
\partial _{\mu }y=\overline{y}_{\mu }/y$ (note that here, at variance with
the situation discussed in App.\ C, partial derivatives with respect to time
are performed keeping constant the position and velocity of the particle).
Similarly we have
\begin{eqnarray}
0 &=&\int_{T}d^{4}x\,\partial _{\mu }J^{\mu }\theta (-x^{0})\theta
(x^{\prime 0})\theta (y-r)  \nonumber \\
&=&\int_{T\cap \pi }d^{3}{\bf x}\,J^{0}\theta (x^{\prime 0})\theta
(y-r)-\int_{T\cap \pi ^{\prime }}d^{3}{\bf x}^{\prime }\,J^{\prime 0}\theta
(-x^{0})\theta (y-r)  \nonumber \\
&&+\int_{\partial T}d\sigma \,J^{\mu }n_{\mu }\theta (-x^{0})\theta
(x^{\prime 0})-\int_{T}d^{4}x\frac{J^{\mu }\overline{y}_{\mu }}{y}\theta
(-x^{0})\theta (x^{\prime 0})\delta (y-r).  \label{D1'}
\end{eqnarray}
By subtracting Eq.\thinspace (\ref{D1'}) to Eq.\thinspace (\ref{D1}) we
obtain
\begin{eqnarray}
&&\int_{T\cap \pi }d^{3}{\bf x}\,J^{0}\theta (y-r)-\int_{T\cap \pi ^{\prime
}}d^{3}{\bf x}^{\prime }\,J^{\prime 0}\theta (y-r)  \nonumber \\
&=&\int_{\partial T}d\sigma \,J^{\mu }n_{\mu }\left[ \theta (x^{0})\theta
(-x^{\prime 0})-\theta (-x^{0})\theta (x^{\prime 0})\right]  \nonumber \\
&&-\int d^{4}x\frac{J^{\mu }\overline{y}_{\mu }}{y}\left[ \theta
(x^{0})\theta (-x^{\prime 0})-\theta (-x^{0})\theta (x^{\prime 0})\right]
\delta (y-r).  \label{D2}
\end{eqnarray}
According to the arguments discussed in Sec.\ II, for $r\rightarrow 0$ the
two terms on the left-hand side of Eq.\ (\ref{D2}) can be expressed in the
form
\begin{equation}
\int_{T\cap \pi }d^{3}{\bf x}\,J^{0}\theta (y-r)=b_{\nu }\left( P^{\nu
}\right) _{-1}r^{-1}+b_{\nu }{\cal R}\int_{T\cap \pi }d^{3}{\bf x}\,T^{0\nu
}+O(r),  \label{D3}
\end{equation}
\begin{equation}
\int_{T\cap \pi ^{\prime }}d^{3}{\bf x}^{\prime }\,J^{\prime 0}\theta
(y-r)=b_{\nu }^{\prime }\left( P^{\prime \nu }\right) _{-1}r^{-1}+b_{\nu
}^{\prime }{\cal R}\int_{T\cap \pi ^{\prime }}d^{3}{\bf x}^{\prime } \,%
T^{\prime 0\nu }+O(r),  \label{D3'}
\end{equation}
where, in accordance with Eqs.\ (\ref{2.12}) and (\ref{2.11}), we have
defined
\begin{eqnarray}
(P^{k})_{-1} &\equiv &\frac{q^{2}}{4\pi }\frac{2}{3}u^{k}  \label{D4} \\
(P^{0})_{-1} &\equiv &\frac{q^{2}}{4\pi }\left( \frac{2}{3}u^{0}-\frac{1}{6}%
\gamma ^{-1}\right)\,,  \label{D4'}
\end{eqnarray}
and similarly for the primed quantities.

Let us now assume that the reference frame $O$ is related to $O^{\prime }$
by a boost transformation bringing the particle instantaneously at rest at
the proper time $\overline{\tau }$. We then have ${\bf v}=0$, $y=\left| {\bf %
x}\right| $, $\overline{y}^{0}=0$, $\overline{y}^{k}=x^{k}$,
$x^{\prime 0}=\gamma^{\prime } \left( x^{0}+{\bf v}^{\prime }
\cdot {\bf x}\right) $, so that the last term on the right-hand
side of Eq.\ (\ref{D2}) can be rewritten as
\begin{equation}
-\int d^{4}x\frac{J^{\mu }\overline{y}_{\mu }}{y}\left[ \theta (x^{0})\theta
(-x^{\prime 0})-\theta (-x^{0})\theta (x^{\prime 0})\right] \delta
(y-r)=r\int_{\left| {\bf x}\right| =r}d\Omega _{{\bf x}}\int_{-{\bf x}\cdot
{\bf v}^{\prime }}^{0}dx^{0}\,x_{i}T^{i\nu }b_{\nu }.  \label{D5}
\end{equation}
Again, it is here useful to expand the energy-momentum tensor as a series of
homogeneous terms of increasing degree in $x$. By applying Eqs.\ (\ref{2.3})
and (\ref{2.5}) with $u^{k}=0$, $u^{0}=1$, one obtains
\begin{eqnarray*}
(T^{i0})_{-4} &=&0 \\
(T^{ij})_{-4} &=&\frac{q^{2}}{(4\pi )^{2}\left| {\bf x}\right| ^{4}}\left(
\frac{1}{2}\delta ^{ij}-\frac{x^{i}x^{j}}{{\bf x}^{2}}\right) \,,
\end{eqnarray*}
so that the $1/r$ term of the expression (\ref{D5}) is given by
\begin{equation}
r\int_{\left| {\bf x}\right| =r}d\Omega _{{\bf x}}\int_{-{\bf x}\cdot {\bf v}%
^{\prime }}^{0}dx^{0}\,x_{i}(T^{i\nu })_{-4}b_{\nu }=-\frac{q^{2}}{8\pi r^{3}%
}\int_{\left| {\bf x}\right| =r}\frac{d\Omega _{{\bf x}}}{4\pi }({\bf x}%
\cdot {\bf v}^{\prime })({\bf x}\cdot {\bf b})=-\frac{q^{2}}{24\pi r}({\bf v}%
^{\prime }\cdot {\bf b}).  \label{D6}
\end{equation}
This result could also have been deduced directly from Eq.\ (\ref{D2}). In
fact, using Eqs.\ (\ref{D3})--(\ref{D4'}) and the relations $b_{\nu
}^{\prime }u^{\prime \nu }=b_{\nu }u^{\nu }=b_{0}$, $b^{\prime 0}=\gamma
^{\prime }(b^{0}+{\bf v}^{\prime }\cdot {\bf b})$, the term of order $1/r$
on the left-hand side of Eq.\ (\ref{D2}) can be written as
\[
b_{\nu }\left( P^{\nu }\right) _{-1}r^{-1}-b_{\nu }^{\prime }\left(
P^{\prime \nu }\right) _{-1}r^{-1}=\frac{q^{2}}{4\pi r}\left[ \frac{2}{3}%
\left( b_{\nu }u^{\nu }-b_{\nu }^{\prime }u^{\prime \nu }\right) -\frac{1}{6}%
\left( b_{0}-\gamma ^{\prime -1}b_{0}^{\prime }\right) \right] =-\frac{q^{2}%
}{24\pi r}({\bf v}^{\prime }\cdot {\bf b}),
\]
in accordance with Eq.\ (\ref{D6}). Thus Eq.\ (\ref{D6}) is strictly related
with the non-covariance of the divergent part $\left( P^{\nu }\right)
_{-1}r^{-1}$ of the non-renormalized electromagnetic energy-momentum.

Now, considering the zeroth order term of Eq.\ (\ref{D5}), we have
\[
r\int_{\left| {\bf x}\right| =r}d\Omega _{{\bf x}}\int_{-{\bf x}\cdot {\bf v}%
^{\prime }}^{0}dx^{0}\,x_{i}(T^{i\nu })_{-3}b_{\nu }=0.
\]
This result follows immediately from the antisymmetry of $(T^{i\nu })_{-3}$
under the change of variables $x^{\mu }\rightarrow -x^{\mu }$. Therefore
from Eqs.\ (\ref{D2})--(\ref{D3'}) we obtain
\begin{equation}
b_{\nu }{\cal R}\int_{T\cap \pi }d^{3}{\bf x}\,T^{0\nu }-b_{\nu }^{\prime }%
{\cal R}\int_{T\cap \pi ^{\prime }}d^{3}{\bf x}^{\prime } \,T^{\prime
0\nu }=\int_{\partial T}d\sigma \,J^{\mu }n_{\mu }\left[ \theta
(x^{0})\theta (-x^{\prime 0})-\theta (-x^{0})\theta (x^{\prime 0})\right] \,.
\label{D7}
\end{equation}

If a single charged particle is present in the whole space, one can consider
the limit of Eq.\ (\ref{D7}) when the spatial dimensions of the tube $T$
tend to infinity. Assuming as usual that $T^{\mu \nu }$ vanishes
sufficiently fast at infinity, the integral over $\partial T$ on the
right-hand side will vanish, so that Eq.\ (\ref{D7}) becomes
\begin{equation}
b_{\nu }P_{{\rm em}}^{\nu }-b_{\nu }^{\prime }P_{{\rm em}}^{\prime \nu
}=b_{\nu }{\cal R}\int_{\pi }d^{3}{\bf x}\,T^{0\nu }-b_{\nu }^{\prime }{\cal %
R}\int_{\pi ^{\prime }}d^{3}{\bf x}^{\prime } \,T^{\prime 0\nu }=0.
\end{equation}
Since $b$ was arbitrarily chosen, the above equation implies that $P_{{\rm %
em}}^{\nu }$ transforms as a
four-vector. One may notice that the particle's equation of motion has not
been used up to now in this section. This means that, in the one-particle
case, the Lorentz covariance of the electromagnetic energy-momentum
four-vector defined by Eq.\ (\ref{Pem}) is a consequence of the Maxwell
equations alone, and is verified at any point of the world-line of the
particle for any arbitrary particle trajectory.

We would like to point out that Eq.\ (\ref{D7}) holds also in the more
general case when in place of $O$ one considers another arbitrary system of
reference $O^{\prime \prime }$ for which ${\bf v}^{\prime \prime }(\overline{%
\tau })\neq 0$. Assuming again that $z^{\prime \prime \mu }(\overline{\tau }%
)=0$, and calling $\pi ^{\prime \prime }$ the space with equation
$x^{\prime \prime 0}=0$, this means that we can write
\begin{equation}
b_{\nu }^{\prime \prime }{\cal R}\int_{T\cap \pi ^{\prime \prime }}d^{3}{\bf %
x}^{\prime \prime } \,T^{\prime \prime 0\nu }-b_{\nu }^{\prime }{\cal R}%
\int_{T\cap \pi ^{\prime }}d^{3}{\bf x}^{\prime } \,T^{\prime 0\nu
}=\int_{\partial T}d\sigma \,J^{\mu }n_{\mu }\left[ \theta (x^{\prime \prime
0})\theta (-x^{\prime 0})-\theta (-x^{\prime \prime 0})\theta (x^{\prime 0})%
\right] \,.  \label{D7'}
\end{equation}

To prove this, one has simply to write down the equivalent of Eq.\ (\ref{D7}%
) with $O^{\prime \prime }$ in place of $O^{\prime }$:
\[
b_{\nu }{\cal R}\int_{T\cap \pi }d^{3}{\bf x}\,T^{0\nu }-b_{\nu }^{\prime
\prime }{\cal R}\int_{T\cap \pi ^{\prime \prime }}d^{3}{\bf x}^{\prime
\prime } \,T^{\prime \prime 0\nu }=\int_{\partial T}d\sigma \,J^{\mu
}n_{\mu }\left[ \theta (x^{0})\theta (-x^{\prime \prime 0})-\theta
(-x^{0})\theta (x^{\prime \prime 0})\right] \,.
\]

Then subtracting this equation from Eq.\ (\ref{D7}) directly yields Eq.\ (%
\ref{D7'}), as it is easy to check. We are going to make use of Eq.\ (\ref
{D7'}) for the analysis of the many-particle case in the following
subsection.

\subsection{The many-particle case}

Let us now assume that $N$ particles are present, with $N>1$.
One can prove that also in this case the total energy-momentum of the system,
defined according to Eq.\ (\ref{2.13}),
transforms as a four-vector when particles and fields
satisfy the equations of motion (\ref{2.18}) and (\ref{2.19})--(\ref{2.22}).
To show this,
let us divide the whole space-time into $N$ tubes $T_{1}$, $T_{2}$,$\ldots $,%
$T_{N}$ each containing the world-line of a single particle (clearly some
tubes will extend spatially to infinity). Given two arbitrary inertial
frames $O$ and $O^{\prime }$, the spaces $\pi $ and $\pi ^{\prime }$ (again
identified by the equations $x^{0}=0$ and $x^{\prime 0}=0$ respectively)
will intersect the world-line of the generic $n$-th particle at two distinct
points corresponding to the values $\tau $ and $\tau ^{\prime }$ of the
proper time. Calling $\overline{x}^{0}\equiv x^{0}(\tau ^{\prime })$, we
introduce also the space $\overline{\pi }$ characterized by the
equation $x^{0}=\overline{x}^{0}$. Repeating the same arguments of App.\ C
we can write
\begin{eqnarray}
&&m_{n}u_{n}^{\mu }(\tau )+{\cal R}\int_{T_{n}\cap \pi }d^{3}{\bf x}\,T^{\mu
0}-m_{n}u_{n}^{\mu }(\tau ^{\prime })-{\cal R}\int_{T_{n}\cap \overline{\pi }%
}d^{3}{\bf x}\,T^{\mu 0}  \nonumber \\
&=&\int_{\partial T_{n}}d\sigma \,n_{\nu }T^{\mu \nu }\left[ \theta
(x^{0})\theta (\overline{x}^{0}-x^{0})-\theta (-x^{0})\theta (x^{0}-%
\overline{x}^{0})\right]  \label{D9}
\end{eqnarray}
(note that the expression in square brackets on the right-hand side is
written in such a way that the above formula holds in both the cases $%
\overline{x}^{0}>0$ and $\overline{x}^{0}<0$). Besides, rewriting Eq.\ (\ref
{D7'}) for the $n$-th particle at the proper time $\tau ^{\prime }$, we have
\begin{eqnarray}
&&b_{\nu }{\cal R}\int_{T_{n}\cap \overline{\pi }}d^{3}{\bf x}\,T^{0\nu
}-b_{\nu }^{\prime }{\cal R}\int_{T_{n}\cap \pi ^{\prime }}d^{3}{\bf x}%
^{\prime } \,T^{\prime 0\nu }  \nonumber \\
&=&\int_{\partial T_{n}}d\sigma \,J^{\mu }n_{\mu }\left[ \theta (x^{0}-%
\overline{x}^{0})\theta (-x^{\prime 0})-\theta (\overline{x}%
^{0}-x^{0})\theta (x^{\prime 0})\right] \,.  \label{D10}
\end{eqnarray}
Multiplying Eq.\ (\ref{D9}) by $b_{\mu }$ and summing the result to Eq.\ (%
\ref{D10}) we arrive at
\begin{eqnarray*}
&&b_{\nu }\left[ m_{n}u_{n}^{\nu }(\tau )+{\cal R}\int_{T_{n}\cap \pi }d^{3}%
{\bf x}\,T^{0\nu }\right] -b_{\nu }^{\prime }\left[ m_{n}u_{n}^{\prime \nu
}(\tau ^{\prime })+{\cal R}\int_{T_{n}\cap \pi ^{\prime }}d^{3}{\bf x}%
^{\prime } \,T^{\prime 0\nu }\right] \\
&=&\int_{\partial T_{n}}d\sigma \,J^{\mu }n_{\mu }\left[ \theta
(x^{0})\theta (-x^{\prime 0})-\theta (-x^{0})\theta (x^{\prime 0})\right] \,.
\end{eqnarray*}
Finally, let us sum over $n$. The integrals over the various $\partial T_{n}$
will cancel out (neglecting as usual the contributions at infinity), so that
we are left with
\[
b_{\nu }P^{\nu }-b_{\nu }^{\prime }P^{\prime \nu }=0,
\]
where the total momenta $P^{\nu }$ and $P^{\prime \nu }$ are defined
according to Eq.\ (\ref{2.13}) in the two reference frames $O$ and $%
O^{\prime }$ respectively. Again, due to the arbitrariness of $b$, the above
equation implies the Lorentz covariance of the total energy-momentum
four-vector.

\section{Variation of the lagrangian}

We shall consider here a single charged particle,
since the generalization of the results to an arbitrary number of
particles is absolutely trivial.
We observe preliminarily that, by differentiating Eq.\ (\ref{3.20})
with respect to $\delta z$, $\delta u$, $\delta \dot{u}$ etc., it
is possible to expand the variation $\delta A$ of the potential in
a similar way:
\begin{equation}
\delta A(z+x)=(\delta A)_{-2}(x)+(\delta A)_{-1}(x)+(\delta
A)_{0}(x)+\delta \widetilde{A}(z)+O(x).  \label{3.23}
\end{equation}
As a function of $x$, $(\delta A)_{j}(x)$ is a homogeneous term of
degree $j$ which under the transformation $x\rightarrow -x$ is odd
for $j=-2$, $j=0$, and even for $j=-1$. Moreover $\delta
\widetilde{A}(z)$ can be expressed as
\begin{equation}
\delta \widetilde{A}^{\mu }\left( z\right) ={\cal
R}\lim_{r\rightarrow 0}\int_{\left| {\bf x}\right|
=r}\frac{d\Omega _{{\bf x}}}{4\pi }\delta A^{\mu }({\bf z+x},t).
\label{3.24}
\end{equation}
We have in particular, using the variables $\overline{y}^{\mu }$
defined by Eq.\ (\ref{A4}),
\begin{equation}
(\delta A^{\mu })_{-2}(x)=\frac{q}{4\pi y^{3}}u^{\mu
}\overline{y}_{i}\delta z^{i}\,.  \label{3.25}
\end{equation}
In expressions such this, it is always assumed that $\delta
z^{i} \ll y$, since in the following we shall first evaluate the
derivatives of $\overline{L}(r)$
for fixed $r$, and only at a later stage we shall consider their
behavior for $r\rightarrow 0$ in order to obtain the corresponding
derivatives of the renormalized lagrangian $L$.

Let us first consider the electromagnetic part of the lagrangian.
Since the fields satisfy Maxwell equations, we can equivalently
replace in Eq.\ (\ref{3.9})
the factor $\theta \left( \left| {\bf x-z}\right| -r\right) $ with $%
\theta (y-r)$, with $y$ defined according to Eq.\ (\ref{2.10}). As
it is
easy to see using Eqs.\ (\ref{2.4})--(\ref{2.6}), Eq.\ (\ref{3.11})
will remain valid, with only a modification in the
coefficient $(L)_{-1}$, which plays no role at all
in our calculations. Using this new definition of $\overline{L}_{{\rm
em}}$\ we have
\begin{equation}
\delta \overline{L}_{{\rm em}}(r)=-\int d^{3}{\bf x}\,F^{\mu \nu
}\partial _{\mu }\delta A_{\nu }\theta (y-r)-\frac{1}{4}\int
d^{3}{\bf x}\,F^{\mu \nu }F_{\mu \nu }\delta y\,\delta (y-r).
\label{3.26}
\end{equation}
Integrating by parts, the first term on the right-hand side can be
rewritten as
\begin{eqnarray}
-\int d^{3}{\bf x}\,F^{\mu \nu }\partial _{\mu }\delta A_{\nu
}\theta (y-r) &=&-\frac{d}{dx^{0}}\int d^{3}{\bf x}\,F^{0j}
\delta A_{j}\theta (y-r) \nonumber \\
&&+\int d^{3}{\bf x}\,\partial _{\mu }F^{\mu \nu
}\delta A_{\nu }\theta (y-r) \nonumber \\
&&+\int d^{3}{\bf x}\,F^{\mu \nu }\delta A_{\nu }\partial _{\mu
}y\,\delta (y-r).  \label{3.27}
\end{eqnarray}
Using the expansions (\ref{2.25}) and (\ref{3.23})
we can write for $r\rightarrow 0$
\begin{equation}
\int d^{3}{\bf x}\,F^{0j}\delta A_{j}\theta (y-r)=\frac{H}{r}
+{\cal R}\int d^{3}{\bf x}\,F^{0j}\delta A_{j}+O(r) \label{contorno}
\end{equation}
where, according to Eqs.\ (\ref{2.5}), (\ref{3.25}) and (\ref{B6}),
\[
\frac{H}{r} =\int d^{3}{\bf x}\,(F^{0j})_{-2}(\delta A_{j})_{-2}\theta
(y-r) =\frac{q^{2}}{12\pi r}u_{i}\delta z^{i}\ .
\]
The second term on the right-hand side of Eq.\ (\ref{3.27}) vanishes
because of Eq.\ (\ref{3.17}), while the third can be
evaluated making use of Eqs.\ (\ref{C2bis}) and (\ref{C3bis}).
After a shift ${\bf %
x-z}\rightarrow {\bf x}$ in the integration variables, followed by
the transformation (\ref{B1}), we obtain
\begin{eqnarray}
\int d^{3}{\bf x}\,F^{\mu \nu }\delta A_{\nu }\partial _{\mu
}y\,\delta (y-r)
&=&\frac{r}{\gamma }\int_{y=r}d\Omega _{{\bf y}}\Big[ F^{i\nu }({\bf z+x}%
)\delta A_{\nu }({\bf z+x})\overline{y}_{i}   \nonumber \\
&& +F^{0i}({\bf z+x})\delta A_{i}({\bf z+x})\left( -\overline{y}%
_{j}v^{j}+x_{j}v^{j}x_{k}\dot{u}^{k}\right) \Big] \,.
\label{3.28}
\end{eqnarray}
We can now substitute into Eq.\ (\ref{3.28}) the expansions
(\ref{2.25}) and (\ref{3.23}), and then discard all those terms
that have vanishing integrals due to their antisymmetry under
space inversion. In this way we obtain
\begin{eqnarray*}
\int d^{3}{\bf x}\,F^{\mu \nu }\delta A_{\nu }\partial _{\mu
}y\,\delta (y-r) &=&\frac{C}{r}+\frac{r}{\gamma }\int_{y=r}d\Omega
_{{\bf y}}\bigg\{ \left[ (F^{i\nu })_{-2}\delta \widetilde{A}_{\nu
}+\widetilde{F}^{i\nu }\left(
\delta A_{\nu }\right) _{-2}\right] \overline{y}_{i}  \\
&& -\left[ (F^{0i})_{-2}\delta \widetilde{A}_{i}+\widetilde{F}%
^{0i}\left( \delta A_{i}\right) _{-2}\right]
\overline{y}_{j}v^{j}\bigg\} +O(r),
\end{eqnarray*}
where $C$ is a term linearly dependent on $\delta z$ and $\delta
u$, which we do not need to evaluate explicitly. The integral on
the right-hand side can be easily calculated by making use of
Eqs.\ (\ref{2.5}), (\ref{3.25}) and (\ref{B6}). This gives
\begin{eqnarray}
&&\int d^{3}{\bf x}\,F^{\mu \nu }\delta A_{\nu }\partial _{\mu
}y\,\delta (y-r)
\nonumber \\
&=&\frac{C}{r}+\frac{q}{\gamma r^{2}}\int_{y=r}\frac{d\Omega _{{\bf y}}}{%
4\pi }\bigg\{ \left[ \left( u^{i}x^{\nu }-u^{\nu }x^{i}\right)
\delta \widetilde{A}_{\nu }+\widetilde{F}^{i\nu }u_{\nu }\left(
x_{j}+x_{k}u^{k}u_{j}\right) \delta z^{j}\right] \overline{y}_{i}
\nonumber \\
&& -\left[ x^{i}\delta
\widetilde{A}_{i}+\widetilde{F}^{0i}u_{i}\left(
x_{h}+x_{k}u^{k}u_{h}\right) \delta z^{h}\gamma ^{-1}\right] \overline{y}%
_{j}u^{j}\bigg\} +O(r)  \nonumber \\
&=&\frac{C}{r}-\frac{q}{\gamma }\left( u^{\nu }\delta \widetilde{A}_{\nu }+%
\frac{1}{3}\widetilde{F}^{\mu i}u_{\mu }\delta z_{i}\right) +O(r).
\label{3.29}
\end{eqnarray}
The second term on the right-hand side of Eq.\ (\ref{3.26}) can be
evaluated with a very similar procedure, using the relation
\[
\delta y=-\frac{1}{y}\left[ x_{i}-z_{i}+\left( x_{j}-z_{j}\right) u^{j}u_{i}%
\right] \delta z^{i}+\frac{1}{y}\left( x_{j}-z_{j}\right)
u^{j}\left( x_{i}-z_{i}\right) \delta u^{i},
\]
which follows from Eq.\ (\ref{2.9}). We obtain
\begin{eqnarray}
-\frac{1}{4}\int d^{3}{\bf x}\,F^{\mu \nu }F_{\mu \nu }\delta
y\,\delta (y-r)
&=&\frac{r}{4\gamma }\int_{y=r}d\Omega _{{\bf y}}F^{\mu \nu }({\bf z+x}%
)F_{\mu \nu }({\bf z+x})\left( \overline{y}_{i}\delta
z^{i}-x_{j}u^{j}x_{i}\delta u^{i}\right)   \nonumber \\
&=&\frac{D}{r}+\frac{r}{2\gamma }\int_{y=r}d\Omega _{{\bf
y}}\left( F^{\mu \nu }\right) _{-2}\widetilde{F}_{\mu \nu
}\overline{y}_{i}\delta z^{i}+O(r)
\nonumber \\
&=&\frac{D}{r}+\frac{q}{\gamma r^{2}}\int_{y=r}\frac{d\Omega _{{\bf y}}}{%
4\pi }u^{\mu }x^{\nu }\widetilde{F}_{\mu \nu
}\overline{y}_{i}\delta
z^{i}+O(r)  \nonumber \\
&=&\frac{D}{r}+\frac{q}{3\gamma }\widetilde{F}^{\mu i}u_{\mu
}\delta z_{i}+O(r).  \label{3.30}
\end{eqnarray}
Therefore, combining Eqs.\ (\ref{contorno}), (\ref{3.29}) and
(\ref{3.30}), we obtain
\begin{equation}
\delta L_{\rm em}={\cal R}\lim_{r\rightarrow 0}
\delta \overline{L}_{{\rm em}}(r)=-\frac{d}{dt}\left(
{\cal R}\int d^{3}{\bf x}\,F^{0j}\delta A_{j}\right)
-q\gamma ^{-1}u^{\nu}
\delta \widetilde{A}_{\nu }\ .  \label{3.31}
\end{equation}

Considering now the interaction lagrangian, we have from Eq.\
(\ref{3.10})
\begin{eqnarray}
\delta \overline{L}_{{\rm int}}(r) &=&q\delta v^{i}\int_{\left| {\bf x}%
\right| =r}\frac{d\Omega _{{\bf x}}}{4\pi }\,A_{i}\left( {\bf z}+{\bf x}%
,t\right) +q\gamma ^{-1}u^{\nu }\int_{\left| {\bf x}\right|
=r}\frac{d\Omega _{{\bf x}}}{4\pi }\,\delta A_{\nu }\left( {\bf
z}+{\bf x},t\right)
\nonumber \\
&&+q\gamma ^{-1}u^{\nu }\int_{\left| {\bf x}\right| =r}\frac{d\Omega _{{\bf x%
}}}{4\pi }\,\partial _{i}A_{\nu }\left( {\bf z}+{\bf x},t\right)
\delta z^{i}\,.  \label{3.32}
\end{eqnarray}
Keeping in mind Eqs.\ (\ref{3.6}), (\ref{3.34}) and (\ref{3.24}) we can
then write
\begin{equation}
\delta L_{\rm int}={\cal R}\lim_{r\rightarrow 0}
\delta \overline{L}_{{\rm int}}(r)=q\delta v^{i}\widetilde{A}%
_{i}(z)+q\gamma ^{-1}u^{\nu }\delta \widetilde{A}_{\nu
}(z)+q\gamma ^{-1}u^{\nu }\partial _{i}\widetilde{A}_{\nu
}(z)\delta z^{i}\,. \label{3.35}
\end{equation}

We have finally
\begin{equation}
\delta L_{\rm part}=mu_i\delta v^i\ . \label{dlpart}
\end{equation}
Therefore, summing Eqs.\ (\ref{3.31}),
(\ref{3.35}) and (\ref{dlpart}), we obtain Eq.\ (\ref{deltal}).

\section{Translational invariance and conservation laws}

Let us consider the Eq.\ (\ref{deltal}), which has been derived
making use only of the field equations.
When also the equation of
motion (\ref{3.39}) for the particles is satisfied, the second
expression in square brackets vanishes and we have (considering a
single particle to simplify the notation)
\begin{equation}
\delta L=\frac{d}{dt}\left[ \left( mu_{i}+q\widetilde{A}_{i}(z)\right)
\delta z^{i}-{\cal R}\int d^{3}{\bf x}\,F^{0j}\delta A_{j}\right] \,.
\label{E1}
\end{equation}

Let us now examine the space translation of the whole system by an
infinitesimal length $d\varepsilon $ along the direction $x^{k}$:
\begin{eqnarray*}
\delta z^{i} &=&\delta _{k}^{i}d\varepsilon , \\
\delta A^{\mu }(x) &=&-\partial _{k}A^{\mu }(x)d\varepsilon .
\end{eqnarray*}
Substituting these expressions in Eq.\ (\ref{E1}), the manifest invariance
of the lagrangian allows us to write
\[
0=\frac{dL}{d\varepsilon }=\frac{d}{dt}\left[ mu_{k}+q\widetilde{A}_{k}(z)+%
{\cal R}\int d^{3}{\bf x}\,F^{0j}\partial _{k}A_{j}\right] =\frac{dP_{k}}{dt}%
\,,
\]
with
\begin{equation}
P_{k}=mu_{k}+q\widetilde{A}_{k}(z)+{\cal R}\int d^{3}{\bf x}\,F^{0j}\partial
_{k}A_{j}\,.  \label{E2}
\end{equation}
We are now going to show that the conserved quantity $P_{k}$ is equal to the
$k$-th component of the total momentum defined by Eqs.\ (\ref{2.13}) and
(\ref{Pem}). We have in fact from Eq.\ (\ref{2.3})
\begin{equation}
{\cal R}\int d^{3}{\bf x}\,F^{0j}\partial_{k}A_{j}=
{\cal R}\int d^{3}{\bf x}\,T^{0k}+{\cal R}%
\int d^{3}{\bf x}\,F^{0j}\partial _{j}A^{k}.  \label{E3}
\end{equation}
But
\begin{eqnarray*}
\int d^{3}{\bf x}\,F^{0j}\partial _{j}A^{\mu }\theta (y-r) &=&-\int d^{3}%
{\bf x}\,F^{0j}A^{\mu }\partial _{j}\theta (y-r)=-\int d^{3}{\bf x}%
\,F^{0j}A^{\mu }\frac{\overline{y}_{j}}{y}\theta (y-r) \\
&=&-\frac{r}{\gamma }\int_{y=r}d\Omega _{{\bf y}}\,\overline{y}%
_{j}(F^{0j})_{-2}\left[ \left( A^{\mu }\right) _{-1}+\widetilde{A}^{\mu }(z)%
\right] +O(r) \\
&=&-\frac{q}{r^{2}} \int_{y=r}\frac{d\Omega _{{\bf y}}}{4\pi }\overline{y}%
_{j}x^{j}\left[ q\frac{u^{\mu }}{4\pi r}+\widetilde{A}^{\mu }(z)\right] +O(r)
\\
&=&q^{2}\frac{u^{\mu }}{4\pi r}-q\widetilde{A}^{\mu }(z)+O(r),
\end{eqnarray*}
so that
\begin{equation}
{\cal R}\int d^{3}{\bf x}\,F^{0j}\partial _{j}A^{\mu }=-q\widetilde{A}^{\mu
}(z).  \label{E4}
\end{equation}
Therefore Eq.\ (\ref{E2}) becomes
\[
P^{k}=mu^{k}+{\cal R}\int d^{3}{\bf x}\,T^{0k},
\]
in accordance with Eqs.\ (\ref{2.13}) and (\ref{Pem}).

Let us now instead consider an infinitesimal
temporal translation of the system's evolution:
\begin{eqnarray*}
\delta z^{i} &=&v^{i}dt, \\
\delta A^{\mu }(x) &=&\partial _{0}A^{\mu }(x)dt.
\end{eqnarray*}
Since the lagrangian $L$ contains no explicit dependence on $t$, we
obtain from Eq.\ (\ref{E1})
\[
\frac{dL}{dt}=\frac{d}{dt}\left[ \left( mu_{i}+q\widetilde{A}_{i}(z)\right)
v^{i}-{\cal R}\int d^{3}{\bf x}\,F^{0j}\partial _{0}A_{j}\right] \,,
\]
or
\[
\frac{dP^{0}}{dt}=0,
\]
with
\begin{eqnarray}
P^{0} &=&\left( mu_{i}+q\widetilde{A}_{i}(z)\right) v^{i}-{\cal R}\int d^{3}%
{\bf x}\,F^{0j}\partial _{0}A_{j}-L  \nonumber \\
&=&mu^{0}+{\cal R}\int d^{3}{\bf x}\,T^{00}+q\widetilde{A}^{0}(z)+{\cal R}%
\int d^{3}{\bf x}\,F^{0j}\partial _{j}A^{0}=mu^{0}+{\cal R}\int d^{3}{\bf x}%
\,T^{00},  \label{E5}
\end{eqnarray}
where the last equality follows again from Eq.\ (\ref{E4}). Therefore $P^{0}$
is equal to the total energy that was introduced in Sec.\ II. In the form
given by Eqs.\ (\ref{E2}) and (\ref{E5}), the four-vector $P^{\mu }$ can
also be immediately identified with the canonical energy-momentum defined in
Sec.\ IV. The derivation of the conservation laws carried out in the present
section clearly represents a generalization of N\"{o}ther's theorem applied to
the lagrangian defined by Eqs.\ (\ref{3.2})--(\ref{3.5}).

\section{Large wave-vector expansion of the field generated by a point
charge}

By inverting Eqs.\ (\ref{5.21})--(\ref{5.22}) one obtains
\begin{equation}
\left( a^{\mu }\right) _{j}({\bf k},t)=\int d^{3}{\bf x}\,e^{-i{\bf k}\cdot
{\bf x}}\left[ k^{0}\left( A^{\mu }\right) _{j}(x)+i\left( \partial
_{0}A^{\mu }\right) _{j-1}(x)\right] \,,  \label{F1}
\end{equation}
where the particle index $n$ has been omitted to simplify the notation.
Therefore Eqs.\ (\ref{5.19})--(\ref{5.20}) can be obtained in a
straightforward way by substituting into Eq.\ (\ref{F1}) the explicit
expressions derived from Eqs.\ (\ref{A7}) and (\ref{A10}), and then carrying
out the Fourier transformation. However, one can obtain Eqs.\ (\ref{5.18})--(%
\ref{5.20}) in a shorter way by considering, for a given time $\bar{t}$, the
field (analogous to the ``Wentzel's field'' of Ref. \cite{due})
\begin{equation}
A_{\bar{t}}^{\mu }(x)\equiv -\int d^{4}y\,D(x-y)j^{\mu }(y)\theta \left(
\bar{t}-y^{0}\right) ,  \label{F2}
\end{equation}
where
\begin{equation}
j^{\mu }(y)=q\frac{u^{\mu }}{\gamma }\delta \left( {\bf y-z}\right)
=q\int_{-\infty }^{+\infty }d\tau \,u^{\mu }(\tau )\delta^{4}
\left( y-z(\tau )\right)  \label{F3}
\end{equation}
is the current density carried by the particle, while $D$ is the usual
invariant function:
\begin{equation}
D(x)=-\frac{i}{(2\pi )^{3}}\int d^{4}k\,e^{ik\cdot x}\delta
(k^{2})\varepsilon (k^{0})=\frac{1}{2\pi }\delta (x^{2})\varepsilon (x^{0}).
\label{F4}
\end{equation}
It is easy to see that $A_{\bar{t}}^{\mu }(x)=A_{{\rm ret}}^{\mu }(x)$ for $%
\left[ x-z\left( \bar{t}\right) \right] ^{2}>0$, so that we have in
particular
\begin{mathletters}
\label{F5}
\begin{eqnarray}
A_{\bar{t}}^{\mu }\left( {\bf x},\bar{t}\right) &=&A_{{\rm ret}}^{\mu
}\left( {\bf x},\bar{t}\right) \\
\partial _{0}A_{\bar{t}}^{\mu }\left( {\bf x},\bar{t}\right) &=&\partial
_{0}A_{{\rm ret}}^{\mu }\left( {\bf x},\bar{t}\right)
\end{eqnarray}
\end{mathletters}
for every ${\bf x}$. From Eqs.\ (\ref{F2})--(\ref{F4}) we get
\begin{eqnarray}
A_{\bar{t}}^{\mu }(x) &=&-q\int d^{4}y\,D(x-y)\int_{-\infty }^{\overline{%
\tau }}d\tau \,u^{\mu }(\tau )\delta^{4} \left( y-z(\tau )\right)
\nonumber \\
&=&\frac{iq}{(2\pi )^{3}}\int d^{4}k\,\delta (k^{2})\varepsilon
(k^{0})\int_{-\infty }^{\overline{\tau }}d\tau \,u^{\mu }(\tau )e^{ik\cdot %
\left[ x-z(\tau )\right] },  \label{F7}
\end{eqnarray}
where $\overline{\tau }$ is the particle's proper time for which $%
z^{0}\left( \overline{\tau }\right) =\bar{t}$. We can write for $\tau $ near
$\overline{\tau }$
\begin{eqnarray*}
z(\tau ) &=&z+us+\frac{1}{2}\dot{u}s^{2}+O(s^{3}) \\
u(\tau ) &=&u+\dot{u}s+O(s^{2}),
\end{eqnarray*}
where on the right-hand sides $z$, $u$ and $\dot{u}$ stand
for $z\left( \overline{\tau }\right) $, $u\left( \overline{\tau }
\right) $\ and $\dot{u}\left( \overline{%
\tau }\right) $ respectively, while $s=\tau -\overline{\tau }$. Substituting
the two above expressions into Eq.\ (\ref{F7}) we obtain
\begin{eqnarray*}
A_{\bar{t}}^{\mu }(x) &=&\frac{iq}{(2\pi )^{3}}\int d^{4}k\,\delta
(k^{2})\varepsilon (k^{0})e^{ik\cdot (x-z)} \\
&&\times \int_{-\infty }^{0}ds\,e^{-ik\cdot us}\left[ 1-ik\cdot
\dot{u}\frac{s^{2}}{2}+O(s^{3})\right]
\left[ u^{\mu }+\dot{u}^{\mu }s+O(s^{2})\right] \\
&=&\frac{iq}{(2\pi )^{3}}\int d^{4}k\,\delta (k^{2})\varepsilon
(k^{0})e^{ik\cdot (x-z)}\left[ i\frac{u^{\mu }}{k\cdot u}+
\frac{\dot{u}^{\mu }}{(k\cdot u)^{2}}-u^{\mu }
\frac{k\cdot \dot{u}}{(k\cdot u)^{3}}+O\left( \left| {\bf k%
}\right| ^{-3}\right) \right] \\
&=&-\frac{q}{2(2\pi )^{3}}\int \frac{d^{3}{\bf k}}{k^{0}}e^{ik\cdot (x-z)}%
\left[ \frac{u^{\mu }}{k\cdot u}+i\frac{(k\cdot \dot{u})u^{\mu }
-(k\cdot u)\dot{u}^{\mu }}
{(k\cdot u)^{3}}+O\left( \left| {\bf k}\right| ^{-3}\right) \right]
+\text{c.c.}
\end{eqnarray*}
Keeping into account Eqs.\ (\ref{F5}), a comparison of the above equation
with Eqs.\ (\ref{5.1}) provides directly Eqs.\ (\ref{5.18})--(\ref{5.20}).
Of course, it is implicitly assumed that the incident field $A_{{\rm in}%
}^{\mu }\equiv A^{\mu }-\sum_{n}A_{n,{\rm ret}}^{\mu }$ is sufficiently
well-behaved, so that its contribution to $a^{\mu }({\bf k},t)$ for large $%
{\bf k}$\ is at most of order $\left| {\bf k}\right| ^{-3}$.

\section{Proof of Theorem 4}

Taking into account Eqs.\ (\ref{4.19}) and (\ref{4.18}), it is clear
that Eqs.\ (\ref{5.13})--(\ref{5.14}) are equivalent to the
equation
\begin{equation}
{\cal R}\int d^{3}{\bf x}\,\Gamma _{{\rm em}}^{\mu }(A,\pi )
=\frac{1}{2(2\pi )^{3}}{\cal R}\int \frac{d^{3}{\bf k}%
}{k^{0}}k^{\mu }a^{\nu }({\bf k},t)a_{\nu }^{\ast }({\bf k},t),  \label{5.17}
\end{equation}
with
\begin{eqnarray}
\Gamma _{{\rm em}}^{0}(A,\pi ) &\equiv &\frac{1}{4}F_{ij}F^{ij}+\frac{1}{2}%
\pi _{\mu }\pi ^{\mu }-\pi ^{i}\partial _{i}A^{0}+\pi ^{0}\partial
_{i}A^{i}
\nonumber \\
&=&\frac{1}{2}\left[ \partial _{0}A_{\mu }\partial _{0}A^{\mu
}+\partial _{i}A_{\mu }\partial _{i}A^{\mu }+\partial _{i}\left(
A^{i}\partial _{j}A^{j}-A^{j}\partial _{j}A^{i}\right) \right] \,,
\label{5.7}
\end{eqnarray}
\begin{equation}
\Gamma _{{\rm em}}^{k}(A,\pi )\equiv -\pi _{\mu }\partial
^{k}A^{\mu }=\partial ^{0}A_{\mu }\partial ^{k}A^{\mu }-\partial
_{j}\left( A^{0}\partial ^{k}A^{j}\right) +\partial ^{k}\left(
A^{0}\partial _{j}A^{j}\right) \,.  \label{5.8}
\end{equation}
Eq.\ (\ref{5.17}) can be rewritten as
\begin{equation}
{\cal R}\lim_{r\rightarrow 0}\Pi_{{\rm em}}^{\mu
}(t,r)={\cal R}\lim_{\Lambda \rightarrow \infty }K_{{\rm em}}^{\mu
}(t,\Lambda )\ , \label{5rlim}
\end{equation}
with
\begin{equation}
\Pi_{{\rm em}}^{\mu }(t,r)\equiv \int d^{3}{\bf x}\,\Gamma _{%
{\rm em}}^{\mu }(A,\pi )\prod_{n=1}^{N}\theta (y_{n}-r)
\label{5.6}
\end{equation}
and
\begin{equation}
K_{{\rm em}}^{\mu }(t,\Lambda )\equiv \frac{1}{2(2\pi )^{3}}\int \frac{d^{3}%
{\bf k}}{k^{0}}k^{\mu }a^{\nu }({\bf k},t)a_{\nu }^{\ast }({\bf
k},t)\theta \left( \Lambda -k^{0}\right)\ .  \label{5.28}
\end{equation}

In order to evaluate the two ${\cal R}$-limits in Eq.\ (\ref{5rlim}), it is
convenient to decompose the field in the following way:
\begin{eqnarray}
\partial _{\nu }A^{\mu } &=&\sum_{n=1}^{N}\left( \partial _{\nu }A_{n}^{\mu
}\right) _{-2}+\left( \partial _{\nu }A^{\mu }\right) _{{\rm res}}
\label{5.23} \\
a^{\mu } &=&\sum_{n=1}^{N}\left( a_{n}^{\mu }\right) _{-1}e^{-i{\bf k}\cdot
{\bf z}_{n}}+\left( a^{\mu }\right) _{{\rm res}}\,.  \label{5.24}
\end{eqnarray}
The two above equations provide the definition for $\left( \partial _{\nu
}A^{\mu }\right) _{{\rm res}}$ and $\left( a^{\mu }\right) _{{\rm res}}$,
which are of course related to each other by an equation similar to (\ref
{5.22}):
\begin{equation}
\left( \partial _{\nu }A^{\mu }\right) _{{\rm res}}(x)=\frac{i}{2(2\pi )^{3}}%
\int \frac{d^{3}{\bf k}}{k^{0}}k_{\nu }\left[ \left( a^{\mu }\right) _{{\rm %
res}}({\bf k},t)e^{i{\bf k}\cdot {\bf x}}-\left( a^{\mu }\right) _{{\rm res}%
}^{\ast }({\bf k},t)e^{-i{\bf k}\cdot {\bf x}}\right] \,.  \label{5.25}
\end{equation}

Let us first consider the expression (\ref{5.28}).
Taking into account Eq.\ (\ref{5.24}), we have for $\Lambda \rightarrow
\infty $
\begin{equation}
K_{{\rm em}}^{\mu }=C^{\mu }\Lambda +D^{\mu }+O(\Lambda ^{-1}),
\label{5.29}
\end{equation}
with
\[
2(2\pi )^{3}C^{\mu }\Lambda =\sum_{n=1}^{N}\int \frac{d^{3}{\bf k}}{%
k^{0}}k^{\mu }\left( a_{n}^{\nu }\right) _{-1}\left( a_{n,\nu }\right)
_{-1}^{\ast }\theta \left( \Lambda -k^{0}\right)
\]
and
\begin{eqnarray}
2(2\pi )^{3}D^{\mu } &=&\sum_{n\neq m}\int \frac{d^{3}{\bf k}}{k^{0}}%
k^{\mu }\left( a_{n}^{\nu }\right) _{-1}\left( a_{m,\nu }\right) _{-1}^{\ast
}e^{-i{\bf k}\cdot \left( {\bf z}_{n}-{\bf z}_{m}\right) }  \nonumber \\
&&+\sum_{n=1}^{N}\int \frac{d^{3}{\bf k}}{k^{0}}k^{\mu }\left[ \left(
a_{n}^{\nu }\right) _{-1}\left( a_{\nu }\right) _{{\rm res}}^{\ast }e^{-i%
{\bf k}\cdot {\bf z}_{n}}+\left( a^{\nu }\right) _{{\rm res}}\left( a_{n,\nu
}\right) _{-1}^{\ast }e^{i{\bf k}\cdot {\bf z}_{n}}\right]  \nonumber \\
&&+\int \frac{d^{3}{\bf k}}{k^{0}}k^{\mu }\left( a^{\nu }\right) _{{\rm res}%
}\left( a_{\nu }\right) _{{\rm res}}^{\ast }\,.  \label{5.30}
\end{eqnarray}
It is easy to check that the integrals appearing in the expression of
$D^{\mu }$ are indeed convergent. One has to observe in this respect
that, for $\left| {\bf k}\right| \rightarrow \infty $,
\[
\left( a^{\nu }\right) _{{\rm res}}({\bf k},t)=\sum_{n=1}^{N}\left(
a_{n}^{\nu }\right) _{0}({\bf k},t)e^{-i{\bf k}\cdot {\bf z}_{n}}+O\left(
\left| {\bf k}\right| ^{-3}\right)
\]
and that, as a result of Eqs.\ (\ref{5.19})--(\ref{5.20}),
\begin{equation}
\left( a_{n}^{\nu }\right) _{-1}\left( a_{n,\nu }\right) _{0}^{\ast }+\left(
a_{n}^{\nu }\right) _{0}\left( a_{n,\nu }\right) _{-1}^{\ast }=0.
\label{5.31}
\end{equation}

Considering now Eq.\ (\ref{5.6}),
we shall explicitly consider only the temporal component,
since the other three can be treated in exactly the same way. We first
observe that the three-divergence on the right-hand side of Eq.\ (\ref{5.7})
gives no contribution to $\Pi_{{\rm em}}^{0}(t,r)$ for
vanishing $r$. In fact,
assuming as usual that the fields vanish sufficiently fast at infinity,
we have
\begin{eqnarray*}
&&\frac{1}{2}\int d^{3}{\bf x}\,\partial _{i}\left( A^{i}\partial
_{j}A^{j}-A^{j}\partial _{j}A^{i}\right) \prod_{n=1}^{N}\theta (y_{n}-r) \\
&=&-\frac{1}{2r}\sum_{n=1}^{N}\int d^{3}{\bf x}\,\left( A^{i}\partial
_{j}A^{j}-A^{j}\partial _{j}A^{i}\right) \overline{y}_{n,i}\delta (y_{n}-r).
\end{eqnarray*}
Each of the $N$ terms of the sum for $n=1,\ldots ,N$ on the right-hand side
of the above equation has the form
\[
-\frac{r}{2\gamma }\int_{y=r}d\Omega _{{\bf y}}\left( A^{i}\partial
_{j}A^{j}-A^{j}\partial _{j}A^{i}\right) \overline{y}_{i}\,,
\]
where the index $n$ has been dropped and a transformation of the space
coordinates of the form (\ref{B1}) has been carried out. Using the
asymptotic expansions (\ref{3.20}) and (\ref{3.33}), with $\left( A^{\mu
}\right) _{-1}=qu^{\mu }/4\pi y$ and $\left( \partial _{\nu }A^{\mu }\right)
_{-2}=q\overline{y}_{\nu }u^{\mu }/4\pi y^{3}$, in the limit $r\rightarrow 0$
the above integral becomes
\begin{eqnarray*}
&&-\frac{r}{2\gamma }\int_{y=r}d\Omega _{{\bf y}}\left\{ \left[ (A^{i})_{-1}+%
\widetilde{A}^{i}\right] \left( \partial _{j}A^{j}\right) _{-2}-\left[
(A^{j})_{-1}+\widetilde{A}^{j}\right] \left( \partial _{j}A^{i}\right)
_{-2}\right\} \overline{y}_{i} \\
&=&\frac{r}{2\gamma }\int_{y=r}d\Omega _{{\bf y}}\frac{q}{4\pi r^{3}}\left\{ %
\left[ \frac{qu^{i}}{4\pi r}+\widetilde{A}^{i}\right] \overline{y}_{j}u^{j}-%
\left[ \frac{qu^{j}}{4\pi r}+\widetilde{A}^{j}\right] \overline{y}%
_{j}u^{i}\right\} \overline{y}_{i}=0.
\end{eqnarray*}

Therefore, by making use of Eq.\ (\ref{5.23}), we obtain
that the behavior of $\Pi_{{\rm em}}^0(t,r)$ for $%
r\rightarrow 0$\ has the form
\begin{equation}
\Pi_{{\rm em}}^0=\frac{C^{\prime 0}}{r}+D^{\prime 0}+O(r)\,, \label{5.26}
\end{equation}
with
\[
\frac{C^{\prime 0}}{r}=\frac{1}{2}\sum_{n=1}^{N}
\int d^{3}{\bf x}\,\left[ \left(
\partial _{0}A_{n,\mu }\right) _{-2}\left( \partial _{0}A_{n}^{\mu }\right)
_{-2}+\left( \partial _{i}A_{n,\mu }\right) _{-2}\left( \partial
_{i}A_{n}^{\mu }\right) _{-2}\right] \theta (y_{n}-r),
\]
\begin{eqnarray}
D^{\prime 0} &=&\sum_{n<m}\int d^{3}{\bf x}\,\left[
\left( \partial _{0}A_{n,\mu
}\right) _{-2}\left( \partial _{0}A_{m}^{\mu }\right) _{-2}+\left( \partial
_{i}A_{n,\mu }\right) _{-2}\left( \partial _{i}A_{m}^{\mu }\right) _{-2}%
\right]  \nonumber \\
&&+\sum_{n=1}^{N}\int d^{3}{\bf x}\,\left[ \left( \partial _{0}A_{\mu
}\right) _{{\rm res}}\left( \partial _{0}A_{n}^{\mu }\right) _{-2}+\left(
\partial _{i}A_{\mu }\right) _{{\rm res}}\left( \partial _{i}A_{n}^{\mu
}\right) _{-2}\right]  \nonumber \\
&&+\frac{1}{2}\int d^{3}{\bf x}\,\left[ \left( \partial _{0}A_{\mu }\right)
_{{\rm res}}\left( \partial _{0}A^{\mu }\right) _{{\rm res}}+\left( \partial
_{i}A_{\mu }\right) _{{\rm res}}\left( \partial _{i}A^{\mu }\right) _{{\rm %
res}}\right]\,.  \label{5.27}
\end{eqnarray}
The fact that $D^{\prime 0}$ is a finite quantity can be as usual verified
using the asymptotic expansion (\ref{3.33}) and observing that
\begin{equation}
\int d^{3}{\bf x}\,\left( \partial _{\nu }A_{n,\mu }\right) _{-2}\left(
\partial _{\lambda }A_{n}^{\mu }\right) _{-1}\theta (y_{n}-r)=0  \label{5.32}
\end{equation}
because of the spatial antisymmetry of the integrand. It may be noted that
Eqs.\ (\ref{5.31}) and (\ref{5.32}) are directly related to each other.
They are respectively the reasons why there
is not a term of order $\ln r$ on the right-hand side of Eq.\ (\ref{5.26})
and a term of order $\ln \Lambda $ on the right-hand side of Eq.\
(\ref{5.29}).

Finally, by substituting Eqs.\ (\ref
{5.22}) and (\ref{5.25}) into Eq.\ (\ref{5.27}), carrying out the
integrations over the ${\bf x}$
space and comparing the result with the right-hand side of Eq.\ (\ref{5.30}%
), it is straightforward to verify that $D^0=D^{\prime 0}$. Keeping
in mind the definition of the ${\cal R}$-limits, the Eq.\ (\ref{5rlim})
is therefore proved.

To complete the demonstration of the theorem, we are now going to prove Eq.\
(\ref{5.15}). We first observe that, according to Eqs.\ (\ref{3.20}) and (%
\ref{3.21}), we have
\[
\widetilde{A}^{\mu }\left( z_{n}\right) =\lim_{x\rightarrow 0}\left[ A^{\mu
}\left( z_{n}+x\right) -\left( A_{n}^{\mu }\right) _{-1}(x)-\left(
A_{n}^{\mu }\right) _{0}(x)\right] \,,
\]
so that, taking into account Eq.\ (\ref{5.21}), we get
\begin{equation}
\widetilde{A}^{\mu }\left( z_{n}\right) =\frac{1}{2(2\pi )^{3}}\int \frac{%
d^{3}{\bf k}}{k^{0}}\left[ \left( a_{n}^{\mu }\right) _{R}({\bf k},t)e^{i%
{\bf k}\cdot {\bf z}_{n}}+\left( a_{n}^{\mu }\right) _{R}^{\ast }({\bf k}%
,t)e^{-i{\bf k}\cdot {\bf z}_{n}}\right] \,,  \label{5.34}
\end{equation}
with
\[
\left( a_{n}^{\mu }\right) _{R}\equiv a^{\mu }-\left( a_{n}^{\mu }\right)
_{-1}e^{-i{\bf k}\cdot {\bf z}_{n}}-\left( a_{n}^{\mu }\right) _{0}e^{-i{\bf %
k}\cdot {\bf z}_{n}}.
\]
On the other hand, we have for $\Lambda \rightarrow \infty $%
\begin{equation}
\frac{1}{2(2\pi )^{3}}\int \frac{d^{3}{\bf k}}{k^{0}}\left[ a_{\mu }({\bf k}%
,t)e^{i{\bf k}\cdot {\bf z}_{n}}+a_{\mu }^{\ast }({\bf k},t)e^{-i{\bf k}%
\cdot {\bf z}_{n}}\right] \theta \left( \Lambda -k^{0}\right) =E_{\mu
}\Lambda +F_{\mu }+O(\Lambda ^{-1}),  \label{5.36}
\end{equation}
with
\begin{eqnarray*}
E^{\mu }\Lambda &=&\frac{1}{2(2\pi )^{3}}\int \frac{d^{3}{\bf k}}{k^{0}}%
\left[ \left( a_{n}^{\mu }\right) _{-1}({\bf k},t)+\left( a_{n}^{\mu
}\right) _{-1}^{\ast }({\bf k},t)\right] \theta \left( \Lambda -k^{0}\right)
\\
&=&\frac{1}{(2\pi )^{3}}\int \frac{d^{3}{\bf k}}{k^{0}}\left( a_{n}^{\mu
}\right) _{-1}({\bf k},t)\theta \left( \Lambda -k^{0}\right) ,
\end{eqnarray*}
\begin{equation}
F^{\mu }=\frac{1}{2(2\pi )^{3}}\int \frac{d^{3}{\bf k}}{k^{0}}\left[ \left(
a_{n}^{\mu }\right) _{R}({\bf k},t)e^{i{\bf k}\cdot {\bf z}_{n}}+\left(
a_{n}^{\mu }\right) _{R}^{\ast }({\bf k},t)e^{-i{\bf k}\cdot {\bf z}_{n}}%
\right] \,.  \label{5.38}
\end{equation}
One has to observe that $\left( a_{n}^{\mu }\right) _{0}$ does not
contribute to Eq.\ (\ref{5.36}) since
\[
\left( a_{n}^{\mu }\right) _{0}({\bf k},t)+\left( a_{n}^{\mu }\right)
_{0}^{\ast }({\bf k},t)=0.
\]
This corresponds to the fact that $\left( A_{n}^{\mu }\right) _{0}$ is an
antisymmetric function of ${\bf x}$, and so does not contribute to the
integral on the right-hand side of Eq.\ (\ref{3.21}). Taking into account
Eq.\ (\ref{5.36}), the comparison of Eq.\ (\ref{5.38}) with Eq.\ (\ref{5.34})
provides finally Eq.\ (\ref{5.15}).

\end{document}